\documentclass[sigconf]{acmart}

\AtBeginDocument{%
  \providecommand\BibTeX{{%
    \normalfont B\kern-0.5em{\scshape i\kern-0.25em b}\kern-0.8em\TeX}}}

\usepackage{algorithm}
\usepackage{algorithmicx}
\usepackage{algpseudocode}
\usepackage{textcase}
\usepackage{subfigure}
\usepackage{hyperref}
\usepackage{amsmath}
\usepackage{amsthm}
\usepackage{algorithm}
\usepackage{caption}
\usepackage{graphicx}
\usepackage{cleveref}
\newtheorem{definition}{Definition}

\begin{document}

\title{Detecting Outlier Patterns with Query-based Artificially Generated Searching Conditions}

\author{Shuo Yu}
\affiliation{%
\department{School of Software}
\institution{Dalian University of Technology}
\city{Dalian}
\postcode{116620}
\country{China}}
\email{y\_shuo@outlook.com}

\author{Feng Xia}
\affiliation{%
\department{School of Engineering, IT and Physical Sciences}
\institution{Federation University Australia}
\city{Ballarat}
\postcode{VIC 3353}
\country{Australia}}
\email{f.xia@ieee.org}

\author{Yuchen Sun}
\affiliation{%
\department{School of Software}
\institution{Dalian University of Technology}
\city{Dalian}
\postcode{116620}
\country{China}}
\email{aircleven@outlook.com}

\author{Tao Tang}
\affiliation{%
\department{School of Software}
\institution{Dalian University of Technology}
\city{Dalian}
\postcode{116620}
\country{China}}

\author{Xiaoran Yan}
\affiliation{%
\department{Network Science Institute}
\institution{Indiana University}
\city{Ballarat}
\postcode{IN 47405-7000}}
\email{e-mail: yan30@iu.edu}

\author{Ivan Lee}
\affiliation{%
\department{School of Information Technology and Mathematical Sciences}
\institution{University of South Australia}
\country{Australia}}
\email{ivan.lee@unisa.edu.au}

\renewcommand{\shortauthors}{Yu, et al.}

\begin{abstract}
In the age of social computing, finding interesting network patterns or motifs is significant and critical for various areas such as decision intelligence, intrusion detection, medical diagnosis, social network analysis, fake news identification, national security, etc. However, sub-graph matching remains a computationally challenging problem, let alone identifying special motifs among them. This is especially the case in large heterogeneous real-world networks. In this work, we propose an efficient solution for discovering and ranking human behavior patterns based on network motifs by exploring a user's query in an intelligent way. Our method takes advantage of the semantics provided by a user's query, which in turn provides the mathematical constraint that is crucial for faster detection. We propose an approach to generate query conditions based on the user's query. In particular, we use meta paths between nodes to define target patterns as well as their similarities, leading to efficient motif discovery and ranking at the same time. The proposed method is examined on a real-world academic network, using different similarity measures between the nodes. The experiment result demonstrates that our method can identify interesting motifs, and is robust to the choice of similarity measures.
\end{abstract}



\keywords{human behaviour, outlier detection, social computing, heterogeneous network, motif}


\maketitle
\section{Introduction}
\label{sec:intro}
\begin{figure}
    \centering
    \includegraphics[width=0.6\linewidth]{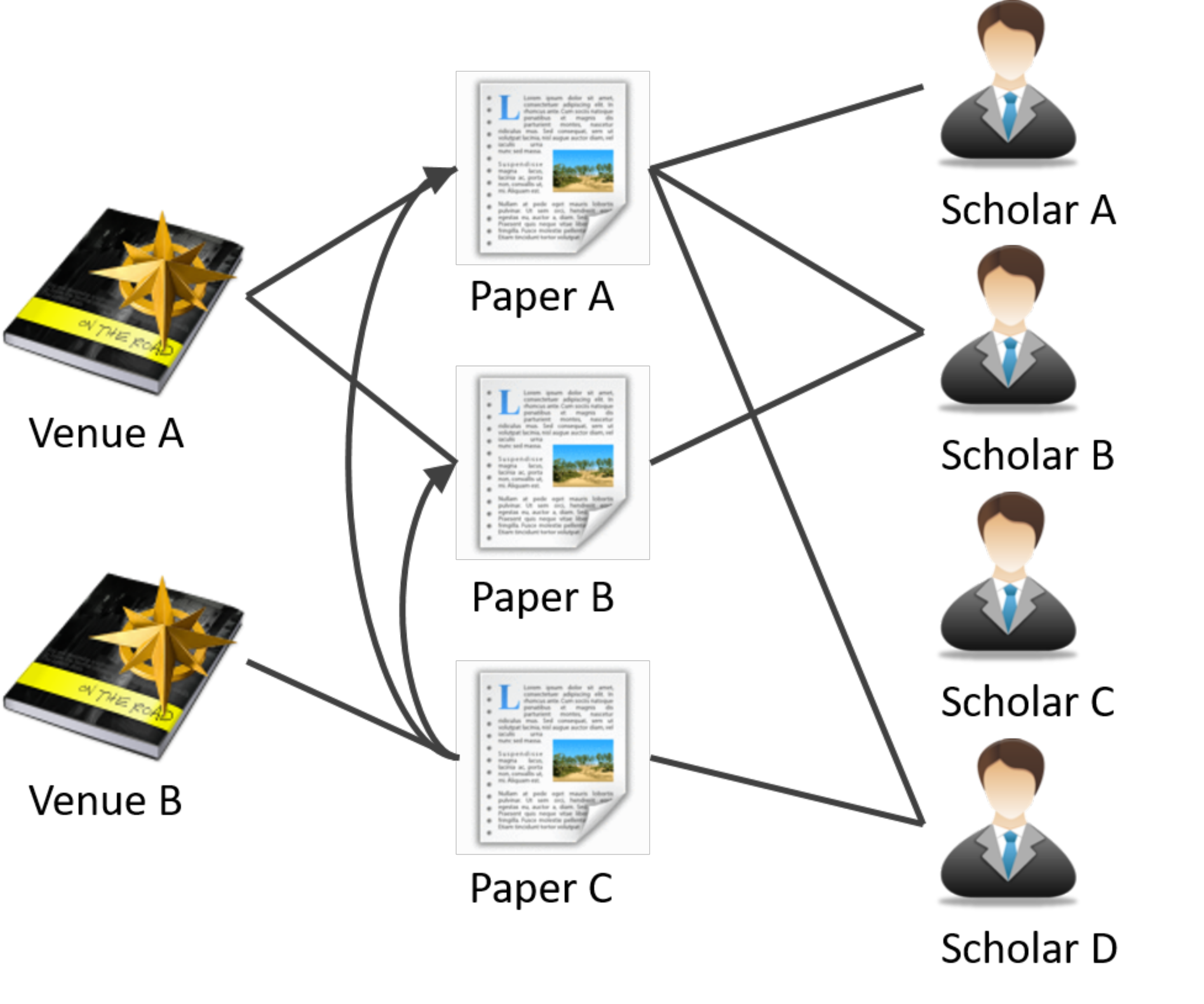}\\
    \caption{An example of a bibliographic network.}\label{fig:hin}
\end{figure}

Networks have been extensively utilized to study the interactions among entities in many fields like social computing. In particular, heterogeneous information networks can model complex, large-scale data sets by introducing multiple node and edge types, leading to more detailed and realistic applications in the real world~\cite{liu2018artificial, Kong2018Mobility, CAIS2017}. Heterogeneous edges can be used to capture relationships among different types of nodes, and edge weights are used to evaluate the strengths of relationships. For example, Figure~\ref{fig:hin} shows a bibliographic network consisting of authors, papers, and venues. In this heterogeneous information network, paper authorship, paper citations, and venue publication relationships are captured at the same time. Additional co-authorship edges can be added to capture collaborative/social relationships between authors, and the collaborative strength can be evaluated by collaboration times~\cite{Yu2017Team}.

In the recent past, lots of effort has been placed into the study of heterogeneous information networks and various relationships among entities. Finding unexpected events and detecting outliers from normal patterns have always been an interesting topic in various disciplines~\cite{lim2018data, chattopadhyay2018scenario}. It can reveal hidden patterns, which may also guide policymakers to make better decisions. These outliers may carry significant information in abundant fields that include, but not limited to, intrusion detection~\cite{kovarasan2019effective}, medical diagnosis~\cite{savadjiev2019demystification}, social network analysis~\cite{kaur2019authorship}, fake news identification~\cite{shu2017fake}, public security monitoring~\cite{zhang2016thermodynamics} and national security~\cite{vaidya2004privacy}. While widely applied to high-dimensional data, uncertain data, streaming data, network data, time-series data, etc., studies about outlier detection mainly focus on individually abnormal node or outlier pairs~\cite{gupta2014outlier,landauer2018time,campos2016evaluation}. Various survey papers summarize the existing outlier detection algorithms from abundant perspectives~\cite{qi2012clustering,yu2018space,domingues2018comparative}.

Beyond node or pairs, outlier detection based on sub-graphs or \textbf{motifs} aims to identify important local structures of networks. Defined by a particular pattern of nodes and edges that is statistically significant, motifs may provide a deep insight into the network's functional building blocks~\cite{milo2002network,shi2017survey}. For example, motifs are widely used in biological networks for the identification of functional DNA sequences and gene regulatory patterns~\cite{georgiev_evidence-ranked_2010,masoudi-nejad_building_2012}. Despite the efforts of scholars from computer science and bioinformatics, motif discovery remains a computationally challenging problem, given its combinatorial nature~\cite{kashtan_efficient_2004,ciriello_review_2008}.

Most motifs discovery algorithms try to enumerate all sub-graph structures in the network and therefore suffer from the exponential complexity as the motif size grows \cite{wernicke2006fanmod}. An alternative approach based on user queries provides more focused and efficient solutions~\cite{speed_network_2007,omidi_moda:_2009}. In the context of heterogeneous information networks, motif queries can be further specified with node and edge types, as well as searching meta paths to enable faster and more precise local detections. For example, many users reviewing one commodity on an online shopping website may be a common pattern. But if many users with similar IDs post the same content in a short period, these users may be of particular interest (e.g., Internet bots \cite{amant2016natural}). The search can be further narrow down by imposing constraints on commodity and user types.

Traditional motif discovery methods rely on statistical null models to find special motifs among the sub-structures and identify outliers~\cite{masoudi-nejad_building_2012,schlauch2015influence}. Here we propose a motif similarity measure based on their meta path connections, which can also be specified in a user's query. For example, in a query of an academic collaboration network, we start with a motif consists of two authors and their co-authored physics paper. If there exist a motif contains two authors and a paper in a discipline that significantly differs from physics (e.g., social science, arts, etc.), it is unlikely that the two motifs will be strongly connected by the query meta path set (author-paper-author and paper-author-paper), especially compared with other multi-authored physics papers. Since the query contains both types of constrained motifs and meta paths, our algorithm can be further optimized for motif discovery and ranking at the same time. Outlier motifs refer to those motifs lying in the searching results but with least similarities comparing to the query motif. For example, we are aiming to find two authors coauthored one paper in physics, and the searching results contain authors coauthored mathematics, chemistry, and art. Then the motif with art may be regarded as an outlier motif because others are all theoretical science.

The high-level procedure of motif discovery is shown in Figure~\ref{fig:framework}. First of all, users should first define one or more target motifs to start the query. These target motifs are regarded as ``motif reference'' when comparing sub-graphs. Then we use meta paths (also specified by the users) to discover sub-graphs that are related to the target motifs. These discovered sub-graphs are ``candidate motifs". After comparing the similarities between candidates and references, the candidate motifs with lower similarities are regarded as outliers or the final output motifs. All of the formal definitions are shown in Section 2.1.

We propose the similarity measure called MOS (Motif Outlier Score) to evaluate the similarity between different candidate motifs. According to user's queries, guided by meta paths, we can efficiently calculate MOS in the candidate motif set based on the statistics gathered during the sub-graph matching process \cite{sun2012efficient}. Finally, the interesting motifs can be recognized according to the ranked scores list. Original contributions in this paper are outlined as follows.

\begin{itemize}
    \item We propose an efficient and flexible motif discovery algorithm by taking advantage of sub-graph and meta path queries (i.e. human behaviour), leading to an intelligent motif searching framework that is applicable to a wide range of social computing applications.
    \item The target motif similarity is defined based on node similarities. Interesting human behavior patterns have been discovered by applying our algorithm to real-world heterogeneous information networks.
    \item We provide empirical evidence that our framework is robust to the choice of similarity measures, including PathSim, CosSim, and MOS.
\end{itemize}

\section{Finding Outlier Motifs Based on Queries}

\label{sec:alg}
Herein, we introduce how to find outlier motifs based on users' queries. We firstly give some related definitions. Then we introduce our proposed outlier detection method. To clearly illustrate our method, we explain it in four steps. First, we achieve motifs based on queries. Second, we count the meta paths. Then, we calculate the similarity of node pairs. Finally, we order the MOS of each motif. When we complete the four steps, we can find the outlier motifs based on queries.

\begin{figure*}[htbp]
    \centering
    \includegraphics[width=0.8\textwidth]{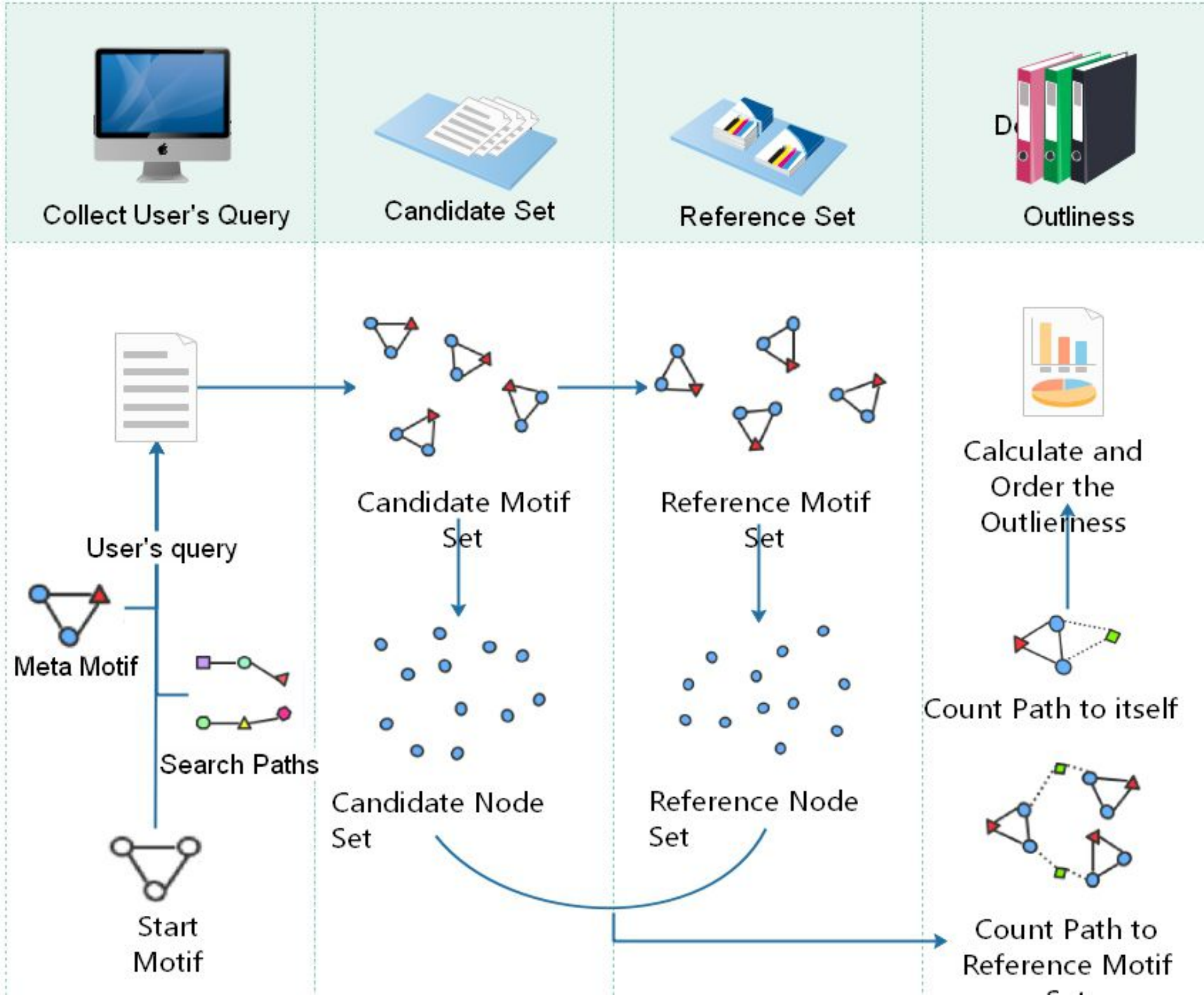}\\
    \caption{The framework of finding interesting outlier motifs based on queries.}\label{fig:framework}
\end{figure*}

As shown in Figure~\ref{fig:framework}, the User's query contains concepts like target motif and search paths and we use an instance to explain them. If an administer wants to find unusual structures of ``two friends that like the same genre of music" around friends of two friends like rock on a streaming media platform. We called entities of the triplet as motif (i.e., Alice and Bob like rock, Alice-Bob-rock is a motif), and target motif is an abstract concept that constrain node types and structures of found motifs. We find their friends by the ``user-user" type paths. Besides, we have to define a standard to evaluate the strange level of these motifs, if the administer wants to compare the motifs around the start motif using motifs generated by another motif (i.e., Cindy and David like blues). The motifs used for comparison form the reference set. We count the number of ``user-genre-user" paths each motif in the candidate set with all motifs in the reference set. If a motif has fewer paths connecting with the reference set, it is more likely to be an outlier.

\subsection{Motifs, Meta path, and Similarity}
\label{sec:mms}
Motifs are generally considered as building blocks in various kinds of networks, including information networks, transportation networks, social networks, and so on~\cite{shi2017survey}. There have been various studies verifying that motifs occupy important positions in networks. Network motifs refer to the small structures frequently appearing in the information network~\cite{Cheng2015A}. Herein, we formally give the definitions of information network, network motif, meta path, and other related concepts.

\begin{definition}
    \label{def:network}
    Information network: An information network can be written as a graph $G(V, E, \varphi)$. $V$ is the node set of the graph. $E$ is the edge set of the graph. $\varphi$ is a mapping function $\varphi : V \rightarrow A$, wherein $A$ is the set of node types.
\end{definition}
In some cases, the mapping function may be $\psi : E \rightarrow R $, where $R$ is the set of edge types. Apparently, the information network $G(V, E, \varphi)$ is homogeneous when $|A| = 1$. If $|A| > 1$, then the information network $G(V, E, \varphi)$ is heterogeneous.

The formal definition of motifs is given in Definition~\ref{def:motif}.

\begin{definition}
    \label{def:motif}
    Network motif: The network motif refers to a subnetwork structure $M\in{G}$, wherein $M$ appears in the network for $k>\beta$ ($\beta>0$) times. Generally, $\beta$ refers to the times that $M$ appears in $G$'s corresponding random network $\widetilde{G}$.
\end{definition}

Motifs appear in a triangle structure in many situations. Some higher-order motifs are also significant but always cause higher computational complexity, especially in large-scale networks~\cite{Li2015Quick}. Properties of nodes and motifs in an information network can be measured in many ways. It is difficult to judge whether a certain motif is an outlier or not without any restrictions. Therefore, in this study, we constrain outlier motifs in the range of a motif set. With the motif set, the scope of finding outlier motifs can be settled. An outlier motif set is generated based on the query motif and the \textbf{meta path}. Herein, the meta path refers to a path connected by different types of nodes. Since there only exists one node type in homogeneous networks, meta path only exists in heterogeneous networks. The formal definition of a meta path is given as follows.

\begin{definition}
    Meta path: $A$ is the node type set. A meta path is a directed path with fixed length and node types in the path. An example of a meta path can be denoted as $ A_{1}\rightarrow A_{2}\rightarrow ...\rightarrow A_{n}$.
\end{definition}

Meta paths in heterogeneous information networks always reflect certain practical meanings. Take the meta path ``user-genre-user" in the music community network as an example, this meta path can be used to describe the users who are interested in the same type of music. Different meta paths represent different practical meanings. If the sequence of node types in a meta path is centrally symmetric, then this meta path is denoted as a symmetric meta path.

Meta path, especially symmetric meta path, can also be applied to measure similarities between different nodes. Generally, two nodes are recognized as similar if they have many meta paths. There are many ways to describe the similarity between two nodes, such as value, topology, etc. Herein, we introduce three kinds of node similarity evaluation metrics, including PathSim~\cite{sun2011pathsim}, CosSim, and Normalized Connectivity~\cite{kuck2015query}. These definitions are shown as follows, respectively.


\begin{definition}
    \label{def:pathsim}
    PathSim: The PathSim of two nodes in the same type is denoted as
    \[S_{PathSim}( x,y) = \dfrac{2\times |\{p_{x\rightarrow y} |p_{x\rightarrow y} \in P\} |}{|\{p_{x\rightarrow x} |p_{x\rightarrow x} \in P\} |+|\{p_{y\rightarrow y} |p_{y\rightarrow y} \in P\} |},\]
\end{definition}
\noindent wherein, \(P\) is the set of fixed symmetric meta paths, and \(p_{x\rightarrow y}\) is one of the symmetric meta paths between $x$ and $y$.

PathSim is used to measure the similarity between two nodes with fixed symmetric meta paths. Besides PathSim, CosSim (cosine similarity) can also be applied to measure the similarity between two nodes. The formal definition is given as follows.
\begin{definition}
    \label{def:cossim}
    CosSim: The cosine similarity of two nodes $x$ and $y$ is defined as
    \[S_{CosSim}( x, y) \ =\ \frac{\sum _{i\in N_{xy}}( |p_{x\rightarrow i} |\times |p_{y\rightarrow i} |)}{||p_{x\rightarrow N_{x}} ||_2\times ||p_{y\rightarrow N_{y}} ||_2},\]
\end{definition}
\noindent wherein, \(N\) is the node set, which is reachable from node \(i\) via meta path \(p\).

Moreover, scholars have also proposed normalized connectivity to evaluate the similarity between two nodes. Normalized connectivity is used to measure the similarity between two nodes with a fixed meta path, as shown in Definition~\ref{def:nc}.
\begin{definition}
    \label{def:nc}
    The normalized connectivity of two objects in the same type is defined as
    \[S_{NorCon}( x, y) = \ \dfrac{|\{p_{x\rightarrow y} |p_{x\rightarrow y} \in P\} |}{|\{p_{x\rightarrow x} |p_{x\rightarrow x} \in P\} |}\]
\end{definition}
\noindent wherein, \(P\) is the set of fixed symmetric meta paths, and \(p_{x\rightarrow y}\) is one of the symmetric meta paths between $x$ and $y$.

\subsection{Problem Definition}
It should be noted that although our algorithm is used for motifs, it can be applied to sub-graphs in different sizes. In this section, we use a sub-graph to represent small graph structures. In other sections, we mainly use motifs.

For a given heterogeneous information network $G=(V,E,\varphi)$ and a user's query sub-graph $M_u=(V_M,E_M,\varphi_M)$, find the candidate sub-graph set $CMS={M_{c_1}, M_{c_2}, \dots, M_{c_n}}$ based on the searching constraint condition $SCC$. Wherein, $SCC$ contains the given meta path set $S_{mp}={MP_1, MP_2, \dots, MP_n}$ and the start nodes for searching. The outlier sub-graph detection problem aims to find an outlier sub-graph set $S_{out}={M_1, M_2, \dots, M_n}$ satisfying that the sub-graph $M_1\in S_{out}$ owns the outlier score $MOS(i)$ that is beyond the default range.

Herein, $M_u$ should contain the complete information, including node types as well as the connections between different nodes. $SCC$ contains two parts, i.e., meta paths and starting nodes. Meta paths determine what the searching routines are, and the starting nodes determine where the search is originated.

\subsection{Outlier Sub-graph}

The measures mentioned above are used to evaluate the similarity between nodes. However, the similarity of sub-graphs still needs to be studied. Herein, we define MOS based on the outlier scores of nodes in the sub-graph, as shown in Definition~\ref{def:outlierness}. MOS is a function of two sets: the reference set $S_R$ and the candidate set $x$. Sub-graphs in the reference set are used as references. Users give reference set as one of the query constraints. It can also be generated based on the query sub-graph. Sub-graphs in the candidate set are those that need to be compared with sub-graphs in the reference set. Therefore, MOS reflects the outlier degree of sub-graphs comparing with the reference set.
\begin{definition}
    \label{def:outlierness}
    MOS: A value to represent the similarity between sub-graphs, which is defined as
    \[\Omega(x,S_{R}) \ =\ \sum _{y\in S_{R}} s(x,y),\]
\end{definition}
\noindent wherein, \(S_{R}\) is the reference set and \(s\) is a well-defined similarity.

\begin{definition}
    \label{def:outliersub-graph}
    Outlier sub-graph: Suppose $M_u$ is a given sub-graph and $M=\{M_1, M_2, \dots, M_n\}$ is the sub-graph set generated based on $M_u$. A sub-graph $M_o$ is regarded as an outlier sub-graph in the information network $G(V, E, \varphi)$, which states that $M_o$ with the value of MOS is an anomaly.
\end{definition}

\subsection{Outlier Sub-graphs Detection}

The overall process of outlier sub-graph detection is shown in Algorithm~\ref{al:omde}. Based on the user's query sub-graph, we first preprocess the data and set the search conditions. Then we generate the candidate sub-graph set and the reference sub-graph set, which are determined by the user's query sub-graph and meta path. Next, we calculate the MOS values of the sub-graphs in the candidate set. Finally, we order the sub-graphs according to their MOS values.

\begin{algorithm}
    \caption{Outlier Sub-graphs Detection}
    \begin{algorithmic}[1]
        \Require Target sub-graph $M$, searching constraint condition $SCC$, meta path set $MP$, reference sub-graph set $RMS$, candidate sub-graph set $CMS$
        \Ensure An ordered list of candidate sub-graphs $Netoutlist$
        \For {meta path $\in$ $MP$} $CMS$.add(search results);
        \EndFor
        \If {$SCC$ = NULL} $RMS = CMS$
        \Else
        \For {meta path $\in$ $MP$} $RMS$.add(search results);
        \EndFor
        \EndIf
        \For {sub-graph $m$ in $CMS$} $Netoutlist$.append()
        \EndFor
        \State sort(MOSlist);
    \end{algorithmic}
    \label{al:omde}
\end{algorithm}

\subsubsection{Generating Requirements Based on Queries}

User query is particularly important that affects the performance of outlier sub-graphs detection. Generally, a detailed query leads to better detection results. In this work, the user query contains two mandatory parts, including at least one sub-graph and a set of meta paths. A sub-graph is judged to be an outlier sub-graph when its MOS value is beyond the reasonable range. As we have illustrated above, the MOS value is calculated based on the candidate set and reference set. Hence the reference set is vital in the whole outlier sub-graph detection process. The reference set can be provided by the user, or it can be generated based on one or more sub-graphs and meta paths given by the user.

\begin{figure*}[htbp]
    \centering
    \includegraphics[width=0.7\textwidth]{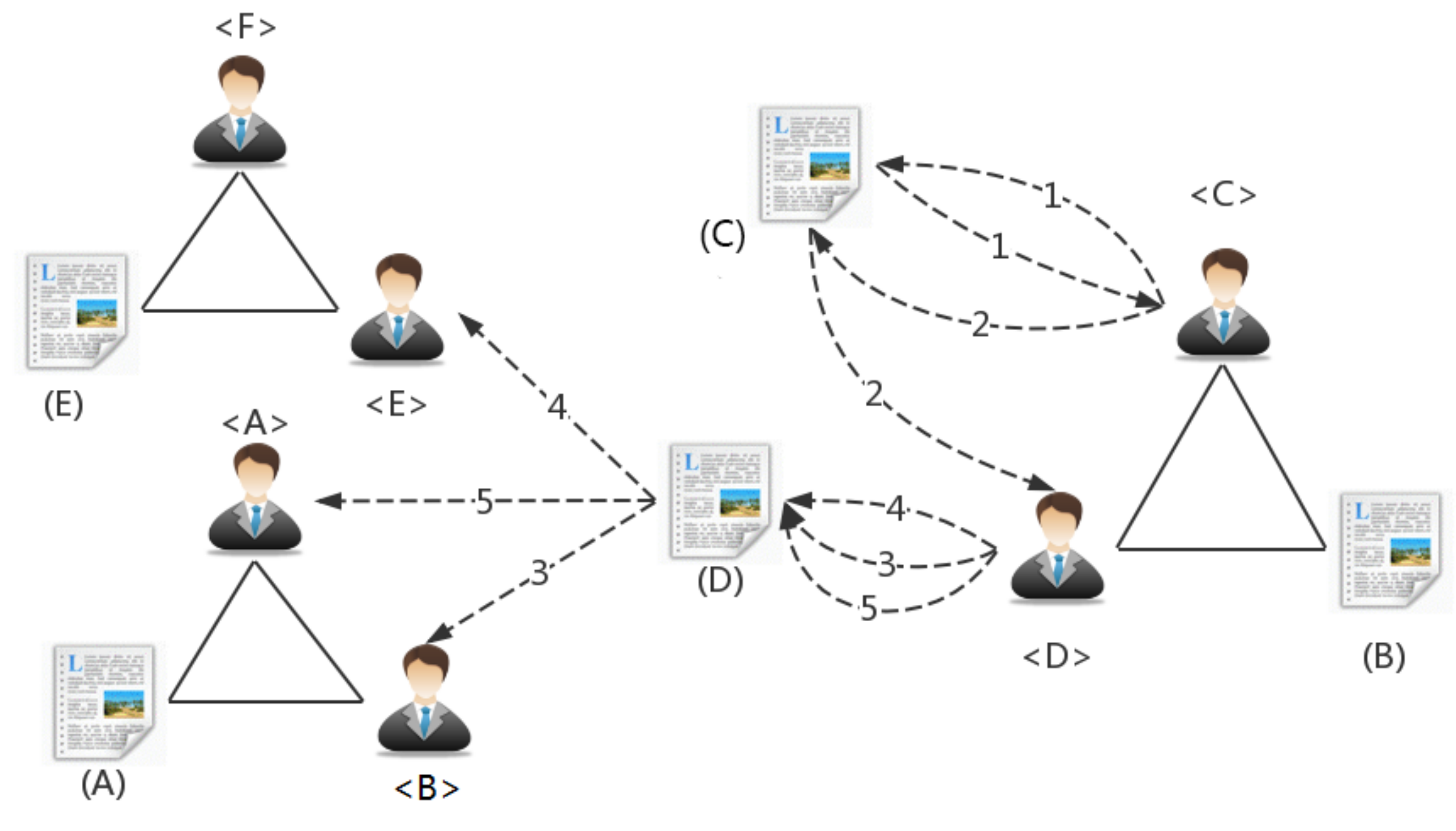}\\
    \caption{A reference set and a candidate set generating process.}\label{fig:exp-req}
\end{figure*}

The generation process of the reference set is the same as that of the candidate set. Figure~\ref{fig:exp-req} shows an example of the process. It is a heterogeneous network that consists of authors and papers. Wherein, the authors are labeled by ``$<>$" and the papers are labeled by ``()". In Figure~\ref{fig:exp-req}, the start sub-graph is ``$<$C$>$-$<$D$>$-(B)" and the type of meta path is given as ``author-paper-author". Then from the start sub-graph, we can find two sub-graphs guide by 5 different meta paths, $<$C$>$-(B)-$<$C$>$, $<$C$>$-(B)-$<$D$>$, $<$D$>$-(D)-$<$B$>$, $<$D$>$-(D)-$<$E$>$ and $<$D$>$-(D)-$<$A$>$. The number labels each path. It can be seen that the 5 meta paths lead us to find three sub-graphs, i.e., $<$F$>$-$<$E$>$-(E), $<$A$>$-$<$B$>$-(A), and $<$C$>$-$<$D$>$-(B). Apparently, $<$C$>$-$<$D$>$-(B) is the start sub-graph itself. Besides, meta path 4 and 5 both lead to the same sub-graph $<$A$>$-$<$B$>$-(A).

With the search conditions of a query given in advance, it is easy to find out the nodes of specified classes. However, there exists a significant problem due to the ordered search. That is, how to avoid the same sub-graph appears many times in the result set?

For every node in a network, we keep the neighbor information classified by node type. We traverse a node by checking whether it is in the map from node instance to positions of meta sub-graph. The first problem can be solved by Depth-First Search (DFS). When generating possible sub-graphs, add the sub-graph into the result set if it (the sub-graph or its isomorphisms) is not found.

\subsubsection{Enumerating Meta Paths}

This procedure is to enumerate the number of meta paths between certain pairs of sub-graphs. As shown in Figure~\ref{fig:exp-req}, there exist two meta paths between sub-graphs $<$A$>$-$<$B$>$-(A), and $<$C$>$-$<$D$>$-(B). In real-world networks, there can be multiple meta paths with a much more complicated situation. To avoid repetitive sub-graph detection, we discover a way to effectively enumerate the meta paths between sub-graphs. That is to enumerate the meta paths between nodes.


Specifically, to get the number of meta paths between any candidate sub-graph and reference set, we first construct a reference node set and a candidate node set, respectively. It should be noted that there could be multiple meta paths used in enumeration, which means that all categories that appeared in the meta sub-graph can be the start node of a meta path. They can also be assigned with different weights. The default weight of each path is 1.

The two nodes connected by a symmetric meta path are generally share higher similarities. This is because a symmetric meta path always connects two nodes of the same type. In this work, we use the number of symmetric meta paths to evaluate the similarity between nodes. For two nodes in a network, the more symmetric meta paths they have, the more similar they are that viewed from different nodes. For two sub-graphs in the network, more symmetric meta paths between nodes at the same positions in sub-graphs, we consider that sub-graphs are more similar in the corresponding positions. If all the nodes at the same positions in two sub-graphs are similar, then we can assume the two sub-graphs are similar. For a given meta path \(MP\), the symmetric meta path of \(MP\) is written as \(MP_{sym} = MP\bigodot MP^{-1}\). The detailed procedures can be seen in Algorithm~\ref{al:frn}. We find the reachable node set for each node in the meta path, and enumerate the number of paths to the reachable node set. After updating the Lastlayer set, Currentlayer is cleared up.

\begin{algorithm}

    \caption{Reachable Nodes Detection}
    \begin{algorithmic}[1]
        \Require Candidate sub-graph set \(CMS\), reference sub-graph set \(RMS\), meta path set $MP$
        \Ensure Attribute node sets $N2N_C$, $N2N_R$
        \Function {GetReachableNodes}{$Sub-graph Set$}
        \State $N2N$ = new dictionary;
        \For {category $c$ in $C$}
        \State initialize Lastlayer;
        \State initialize Currentlayer;
        \For {node $n$ in each sub-graph in $MS$}
        \If {$n$ in N2N[$c$]}
        \State continue
        \EndIf
        \State Lastlayer[$n$] = 1;
        \For {node type $t$ in $mp$}
        \State find the reachable node set from node $n$;
        \State update Currentlayer with paths numbers;
        \State Lastlayer = Currentlayer;
        \State initialize Currentlayer;
        \EndFor
        \State $N2N$[$c$][$n$] = Lastlayer;
        \EndFor
        \EndFor
        \State \Return{$N2N$}
        \EndFunction
        \State initialize category set $C$ = new set;
        \For {meta path $mp$ in $MP$}
        \State $C$.add(state node category in $mp$)
        \EndFor
        \State $N2N_C$=GetReachableNodes($CMS$)
        \State $N2N_R$=GetReachableNodes($RMS$)
        \State \Return $N2N_C$, $N2N_R$;
    \end{algorithmic}
    \label{al:frn}
\end{algorithm}

Herein, we propose a novel search procedure called bidirectional search. Generally, the default search process is a directed search procedure. However, for a symmetric meta path we employed in this work, we use bidirectional search to reduce the complexity. We implement our search process beginning from both the candidate node set and the reference node set. Since the search process is bidirectional, we only search for the half-length of the meta path. This can ensure that the ending nodes are with the same node type.


\subsubsection{Calculating Similarity}

We have introduced three ways to calculate the similarity between two nodes. However, each of them cannot be applied to calculate the sub-graphs similarity directly. Since a sub-graph may have nodes in the same categories in processing, meta paths may start from a sub-graph and return to itself. To be specific, for node similarity calculation, we only count meta paths from a node to itself. Nevertheless, for sub-graph similarity calculation, some sub-graphs may contain nodes in the same category. It leads to a situation that a symmetric meta path from a node can arrive at other nodes with the same node type. The example shown in Figure~\ref{fig:exp-req} has already reflected this kind of situation. Meta paths 1 and 2 both start from the start node, i.e., scholar $<$C$>$. However, meta path 1 ends with the start node $<$C$>$ and meta path 2 ends with scholar $<$D$>$. Though $<$C$>$ and $<$D$>$ are with the same type, $<$C$>$ is apparently not what we want to search for.

As a result, the number of meta paths can be represented as shown in equation \ref{e1}
\begin{equation}
\label{e1}
|\{p_{m\rightarrow m} |p_{m\rightarrow m} \in P\} |\ =\ \sum _{x\in C} |\{p_{x\rightarrow x}\} |\ +\ \sum _{x,y\in C} |\{p_{x\rightarrow y}\} |
\end{equation}
in which \(C\) is the category of \(x\) and \(y\), \(\ x\neq y\).


Before we calculate MOS values of candidate sub-graphs, we have to compute normalized connectivity between nodes. The process for calculating similarities is shown in Algorithm~\ref{al:cmp}.

\begin{algorithm}

    \caption{Meta Paths Calculation}
    \begin{algorithmic}[1]
        \Require $N2N_C$, $N2N_R$, node type set $C$
        \Ensure two maps $A2A, A2B$
        \State initialize $A2A, A2B$ = new set;
        \For {node $c$ in node type set $C$}
        \For {node pair ($a$,$b$) in $N2N_C[c]$}
        \If {\(N2N_{C}[c][a] \cap N2N_{R}[c][b] \neq \emptyset \) and \((a,b) \notin A2B\)}
        \State \(A2B[a][b]=0\);
        \For {$d \in N2N_{C}[c][a] \cap N2N_{R}[c][b]$}
        \State \(A2B[a][b] += N2N_{C}[c][a][d] \times N2N_{R}[c][b][d]\);
        \EndFor
        \EndIf
        \EndFor
        \For {node $e$ in $N2N_C[c]$}
        \If {$e$ in $N2N_C[c]$}
        \If {($a,e$)$\notin A2A$}
        \State $A2A[a][e] = 0$;
        \EndIf
        \For {$d \in N2N_{C}[c][a] \cap N2N_{R}[c][e]$}
        \State\(A2A[a][e] += N2N_{C}[c][a][d] \times N2N_{C}[c][e][d]\);
        \EndFor
        \EndIf
        \EndFor
        \EndFor

        \State \Return $A2A,A2B$
    \end{algorithmic}
    \label{al:cmp}
\end{algorithm}

\subsubsection{Ordering MOS}

In the last step, we calculate MOS of each sub-graph in the candidate set, then we order them according to their MOS values. The result is an ordered list consisting of all sub-graphs in the candidate set. If the MOS of a sub-graph is low, the sub-graph is more likely to be the outlier pattern in the network.

\subsection{Complexity Analysis}

The complexity analysis can be divided into three steps. For time complexity, we firstly consider a situation that the graph is complete. There exist $c$ types of nodes, $c$ is a constant. The number of nodes in the graph is $n$, the number of nodes in each type is $n/c$. We define the length of search path as $l_{1}$ and the size of meta sub-graph as $s$. The time complexity of the first step (1) finding the candidate set of sub-graphs is $O((n/c)^{(l_{1} - 1)} \times (n/c)^{s}) = O(n^{(l_{1} + s -1)})$. In the second step (2), assume the size of the candidate set is $m$ and the length of half of the given symmetric path is $l_{2}$. We use dictionaries to save found nodes and number of paths to them, and the complexity of counting edges is $O((n/c) \times (n/c)^{l_{2} - 1} + m ^ {2}) = O(n^{l_{2}} + m^{2})$. We sort sub-graphs according to their MOS in the last step (3) , whose complexity is $O(m\log(m))$.

However, networks, in reality, are often sparse. As a result, we use and average node degree to compute the complexity of our method. Assume the average node degree is $k, \text{ and } k \ll n$, then the time complexity of the first step (1) is $O(k^{(l_{1} - 1)} \times k^{s}) = O(k^{(l_{1} + s)})$. Assume the nodes of candidate set are $n_{c}$, the complexity of the second step (2) is $O(n_{c} \times k^{(l_{2}-1)} + m^{2}) = O(n_{c} \times k^{l_{2}} + m^{2})$. The last step (3), we sort the similarities of sub-graphs in the candidate set, the complexity in this step is $O(m\log(m))$. We list the two complexity in Table \ref{tab:cmp}.
\begin{table}[h]
    \centering
    \caption{Comparison of the complexity of two types of networks.}
    \resizebox{\linewidth}{!}{
        \begin{tabular}{|c|c|c|}
            \hline
            & Complete graph & Average degree=k graph \\
            \hline
            (1) &$O(n^{(l_{1} + s -1)})$&$O(k^{(l_{1} + s -1)})$\\
            \hline
            (2) &$O(n^{l_{2}} + m^{2})$&$O(n_{c} \times k^{l_{2}} + m^{2})$\\
            \hline
            (3) &{$O(m\log(m))$}&{$O(m\log(m))$}\\
            \hline
        \end{tabular}
    }
    \label{tab:cmp}
\end{table}

\section{Interesting Outlier Motifs in Real-World Data Sets}
\label{sec:exp}
In this section, to verify the effectiveness of the proposed method, we employ it in a real-world data set. Herein, we mainly apply the method in academic networks.

\subsection{Data Set and Experimental Settings}
We employ part of data in Aminer\footnote{\url{http://arnetminer.org/lab-datasets/aminerdataset}} to construct a heterogeneous information network. It contains information including index, title, venue, and other terms of paper. However, not all papers have complete information. We extract 2,092,356 papers and 1,571,933 authors.

When we construct this heterogeneous network, we use three node types, including venues, authors, and terms. The node type of ``paper" is not involved in the construction of the heterogeneous network. This is because the node type ``paper" is the indirect relationship between authors and terms. Terms are extracted from titles of paper, and authors are connected with certain terms when titles of their published papers contain these terms. The types of edges contain ``author - author"  (co-authorship), ``author - term" (via a paper), ``term - term" (appearing in the same title), ``author - venue" (publication), and ``term - venue" (publication).

%

Interestingly, by the above mentioned preprocessing procedure, nodes with super high degree are generally irrelevant to our target. For example, the venue node with the largest degree is ``IEEE Transactions on Information Theory". The degree is 11,227. If we search motifs consisting of this node, the complexity of the search is more than \(\binom {10000} {2} = 5 \times 10^{7}\). The term node with the largest degree is ``of", which appears for 698,767 times. Although high degree nodes in the network mean high frequency, these nodes are insignificant, especially when we want to distinguish outlier motifs. The common feature of motifs is obvious, but finding out underlying relations without any help is a difficult task. In terms, words with high frequency are some usual words like ``of", ``in", ``the", etc. Therefore, we preprocess them by threshold filtering, and those appear for more than 5,000 times are filtered.

\subsubsection{Case 1}
We choose some typical queries to analyze experimental results. First, we use ``F. Xia (author) - G. Wu (author) - authentication (term)" from ``Mobile Networks and Applications" as query motif, ``author - author - term" as target motif, the motif set searched by ``author - term - author" as candidate set and reference set, $\{$``author - term - author", ``term - author - term"$\}$ as the meta path set for MOS calculation. The result set consists of 11,219 motifs. We show the top 10 motifs, which are most similar to the motif set and least similar to the motif set in Table \ref{tab:ex1}.

\begin{table}[h]
    \centering
    \caption{The top 10 outlier motifs and the top 10 most similar motifs detected based on the query motif of ``F. Xia (author) - G. Wu (author) - authentication (term)".}
    \resizebox{\linewidth}{!}{
        \begin{tabular}{|c|c|c|c|}
            \hline
            Author 1 & Author 2 & Term & MOS\\
            \hline
            X. Geng & Y. Huang & defending & 58.867\\
            X. Geng & Y. Huang & against & 58.867\\
            X. Geng & Y. Huang & challenge & 58.867\\
            X. Geng & Y. Huang & ddos & 58.867\\
            X. Geng & A. B. Whinston & defending &  58.867\\
            X. Geng & A. B. Whinston & against & 58.867\\
            X. Geng & A. B. Whinston & challenge & 58.867\\
            X. Geng & A. B. Whinston & ddos &  58.867\\
            Y. Huang & A. B. Whinston & defending & 58.867\\
            Y. Huang & A. B. Whinston & against & 58.867\\
            \(\vdots\) & \(\vdots\) & \(\vdots\) & \(\vdots\)\\
            Q. Zhang & K. Farkas & more & 1817.762\\
            M. Krunz & M. Z. Siam & beaconless &  1848.091\\
            M. Krunz & M. Z. Siam & geographical &  1848.091\\
            Q. Zhang & J. Liu & more & 1855.639\\
            C. Politis & T. Dagiuklas & extreme &  1866.679\\
            H. Chen & C. Yeh & 802.16 & 1868.442\\
            H. Chen & T. Wang & 802.16 & 1868.442\\
            M. Krunz & M. Z. Siam & gains & 1883.091\\
            Q. Zhang & X. Wang & more & 1899.95\\
            Y. Chen & Y. Huang & advanced & 1931.042\\
            \hline
        \end{tabular}
    }
    \label{tab:ex1}
\end{table}

\begin{figure}[!htp]
    \centering
    \subfigure[Terms]{
        \includegraphics[width=0.2\textwidth]{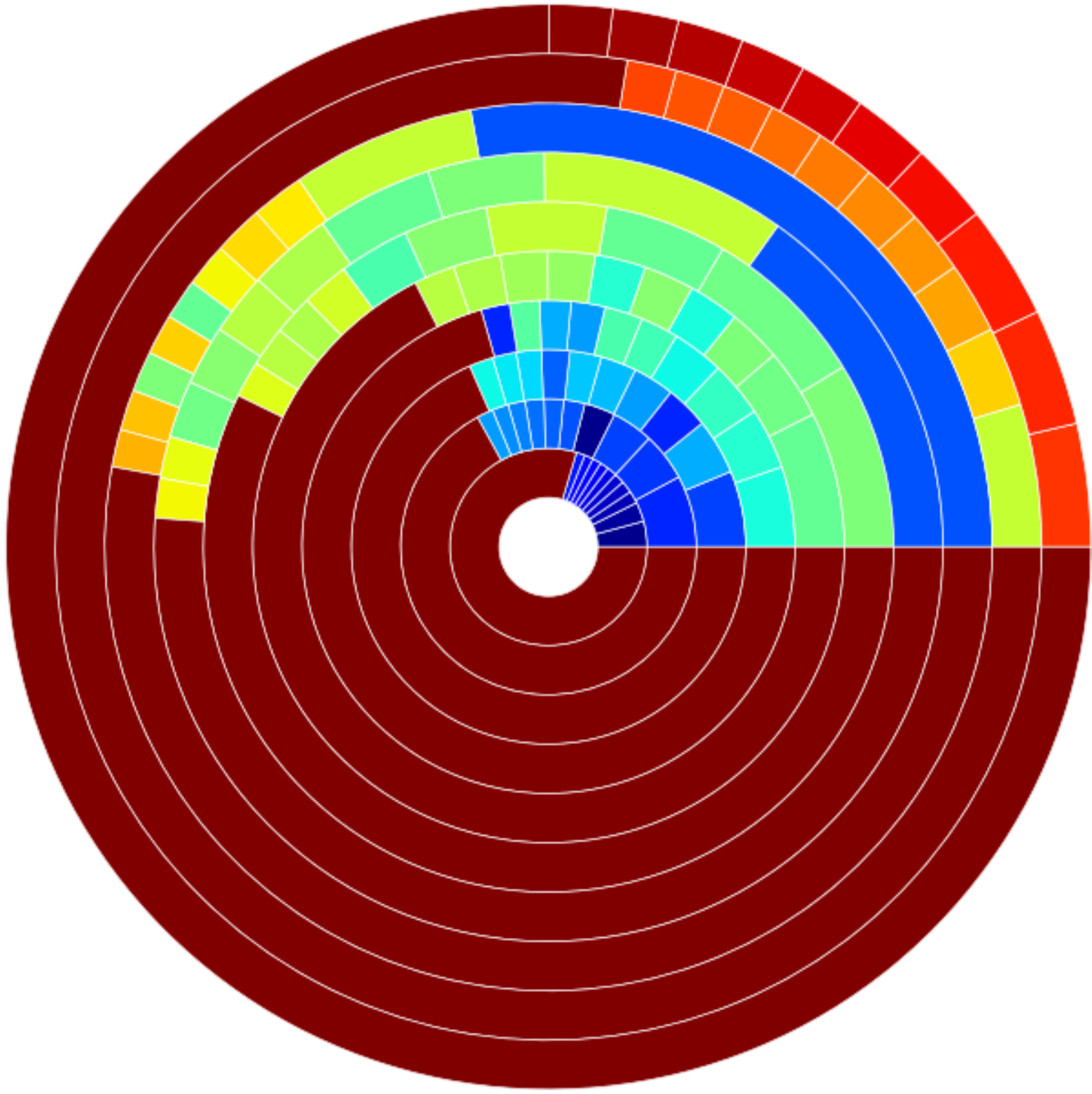}
        \label{fig:perc1_k}
    }
    \subfigure[Authors]{
        \includegraphics[width=0.2\textwidth]{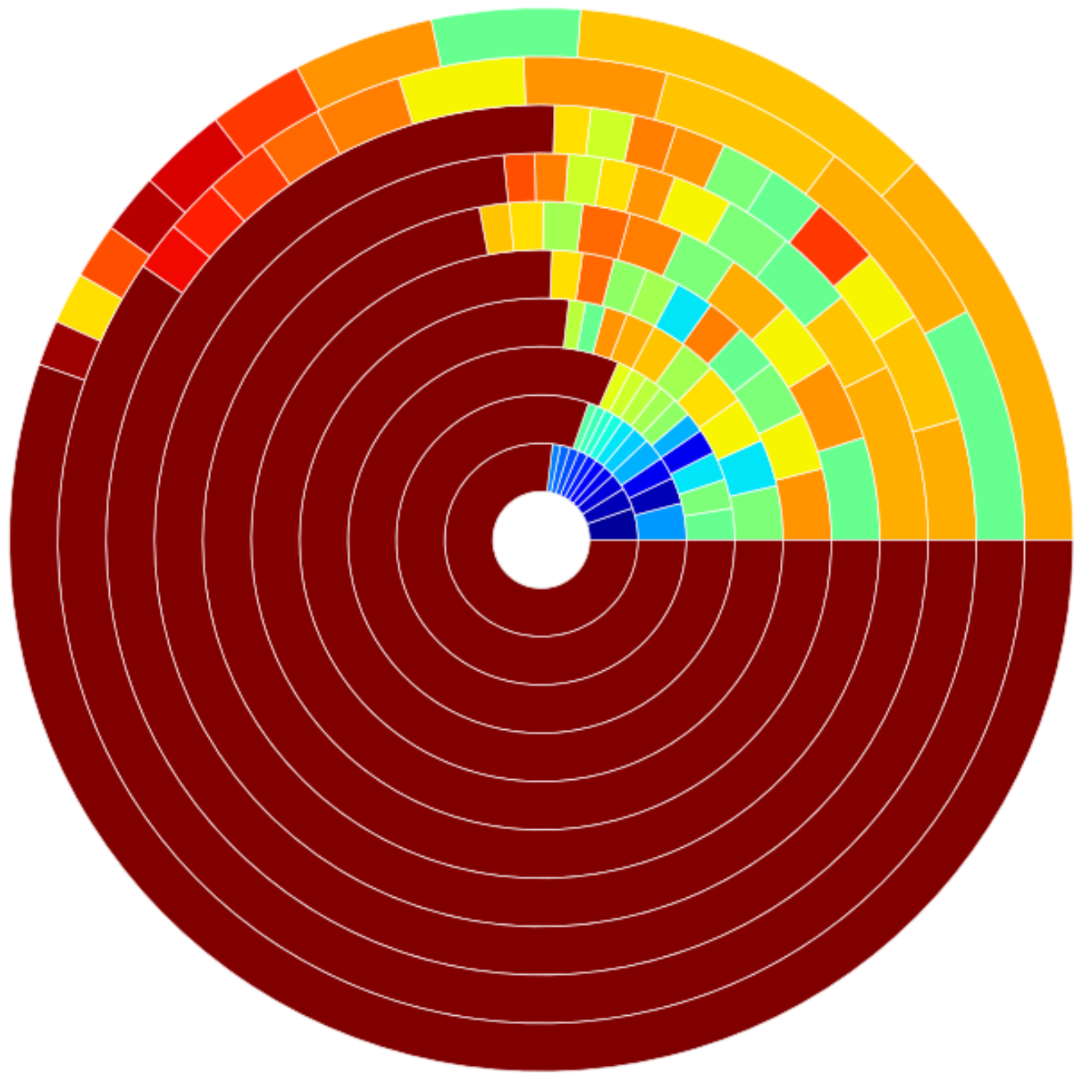}
        \label{fig:perc1_a}
    }
    \caption{The distribution of nodes in outlier motifs. \ref{fig:perc1_k} the distribution of ``terms". \ref{fig:perc1_a} the distribution of ``authors".}
    \label{fig:perc1}
\end{figure}

Although irrelevant words are removed, we still find some useful information - ``802.16" is more relative to ``authentication" than ``ddos". The more interesting thing is that if the motifs are extracted from the same paper, the MOS of motifs tends to be similar - they are not the same, like a motif appearing in two papers. We give more information in Figure~\ref{fig:perc1}.

We divide the motif list into 10 groups according to MOS values. Figure~\ref{fig:perc1_k} shows the distribution of the top 10 outlier terms appearing in each group. Figure~\ref{fig:perc1_a} shows the distribution of the top 10 outlier authors appearing in each group.

To be specific, the different color stands for the MOS value between motifs. The more the color appear in blue, the corresponding term is closer to the outlier. In the old wireless network standard, 802.11 is less relative than 802.16. Research in 802.16 is highly associated to authentication. Moreover, ``authentication" is used frequently in network security, so we can see terms like privacy are highly related to the motif set. If we take network security as the central semantic of the motif set, it is obvious the MOS represents domains related to the central semantic, while others are components of the network.

\subsubsection{Case 2}
In this case, the given query venue is ``IEEE Transactions on Knowledge and Data Engineering". We use ``J. Tang (author) - J. Li (author) - recommendation (term)" as a query motif and ``author - author - term" as target motif type. The motif set searched by ``author - term - author" is regarded as the candidate set and the reference set. We use a set of {``author - term - author", ``term - author - term"} as the meta path set for MOS calculation. The result consists of 26,782 motifs. We list 20 motifs in Table \ref{tab:ex2}. The 20 motifs include 10 motifs, which are most similar to the motif set and another 10 motifs, which are least similar to the motif set.

\begin{table}[h]
    \centering
    \caption{The top 10 outlier motifs and top 10 most similar motifs detected based on the query motif of ``J. Tang (author) - J. Li (author) - recommendation (term)".}
    \resizebox{\linewidth}{!}{
        \begin{tabular}{|c|c|c|c|}
            \hline
            Author 1 & Author 2 & Term & MOS\\
            \hline
            W. Li&D. yeung&mild&34.75\\
            W. Li&D. yeung&multiple-instance&34.75\\
            M. G. Hwang&C. Choi&sense&63.857\\
            M. G. Hwang&P. Kim&sense&63.857\\
            C. Choi&P. Kim&sense&63.857\\
            M. G. Hwang&C. Choi&enrichment&68.538\\
            M. G. Hwang&P. Kim&enrichment&68.538\\
            C. Choi&P. Kim&enrichment&68.538\\
            Z. Chen&C. Wu&component&84.429\\
            A. N. Zincir-Heywood&M. I. Heywood&platforms&85.615\\
            \(\vdots\) & \(\vdots\) & \(\vdots\) & \(\vdots\)\\
            W. Zhang&L. Brankovic&borda&7306.0625\\
            K. Yi&J. Jestes&nutshell&7389.433\\
            K. Yi&J. Jestes&concise&7389.433\\
            K. Yi&D. Srivastava&nutshell&7395.054\\
            K. Yi&D. Srivastava&concise&7395.054\\
            W. Zhang&J. Wang&count&7417.208\\
            X. Lian&K. Yi&constant&7436.015\\
            W. Zhang&J. Wang&consensus-based&7726.739\\
            W. Zhang&J. Wang&multivalued&7726.739\\
            W. Zhang&J. Wang&borda&7726.739\\
            \hline
        \end{tabular}
    }
    \label{tab:ex2}
\end{table}

\begin{figure}[!htp]
    \centering
    \subfigure[Terms]{
        \includegraphics[width=0.2\textwidth]{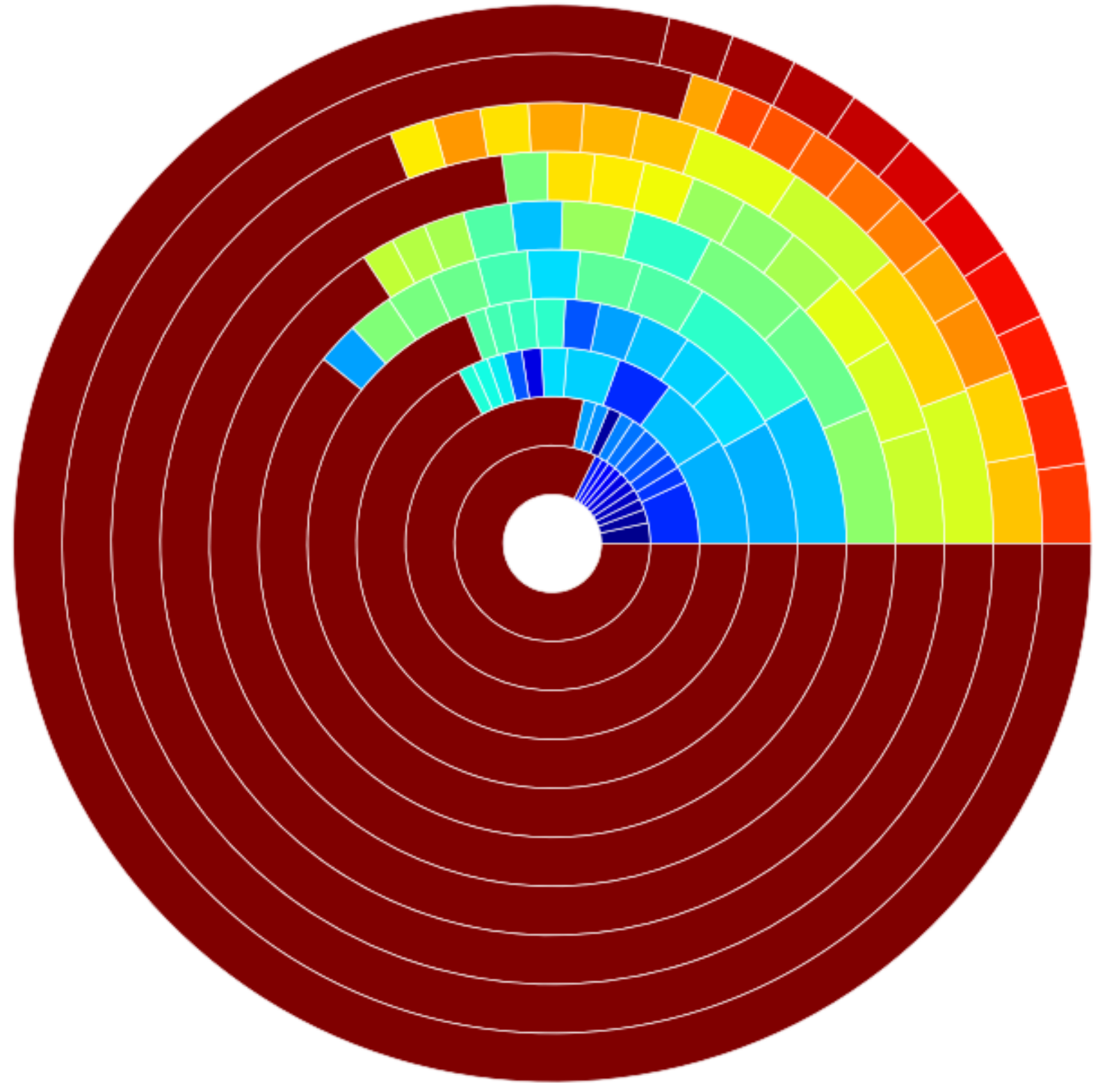}
        \label{fig:perc2_k}
    }
    \subfigure[Authors]{
        \includegraphics[width=0.2\textwidth]{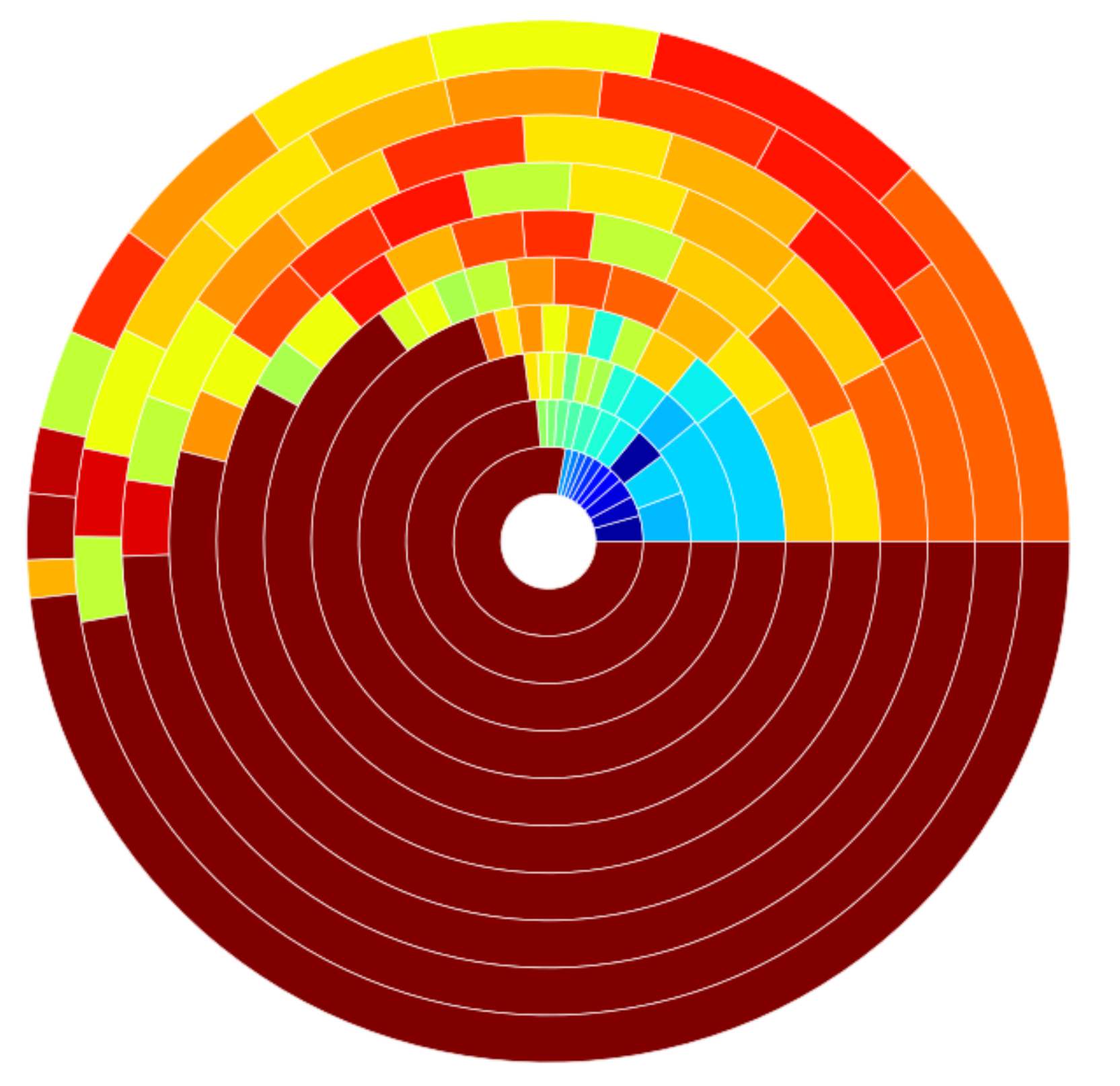}
        \label{fig:perc2_a}
    }
    \caption{The distribution of nodes in outlier motifs. \ref{fig:perc2_k} the distribution of ``term". \ref{fig:perc2_a} the distribution of ``authors".}
    \label{fig:perc2}
\end{figure}

The top 10 outliers is almost irrelevant to the central semantics of the motif set. But different from the top 10 outliers, the top 10 similar motifs are not closely related to our theme of the query motif. The distribution of outlier nodes in this experiment are shown in Figure \ref{fig:perc2_k} and Figure \ref{fig:perc2_a}, respectively.

Most of the words appearing in Figure \ref{fig:perc2_k} are related to recommendation. However, it seems that recommendation is not related to the motif sets. If we focus on the distribution of motifs containing recommendation in the result, there only exists a few parts in the result. Another surprising conclusion is that the nodes with excessive degree will affect our result since they will generate more motifs. Despite the simple preprocess, the results of the experiment are ambiguous. In the next section, we will use a well-defined data set to verify our method and assess its performance.


\section{Discussion}
\label{sec:dis}
Herein, we discuss several critical problems that are crucial to experimental results. First, we show the efficiency of our algorithm under different conditions. Second, we discuss how to choose a candidate set. Then, we discuss how the meta paths affect outlier motif detection. Finally, we discuss the experimental results when using different similarities to detect outlier motifs.

\subsection{Efficiency under different conditions}
\begin{figure}
    \centering
    \includegraphics[width=0.8\linewidth]{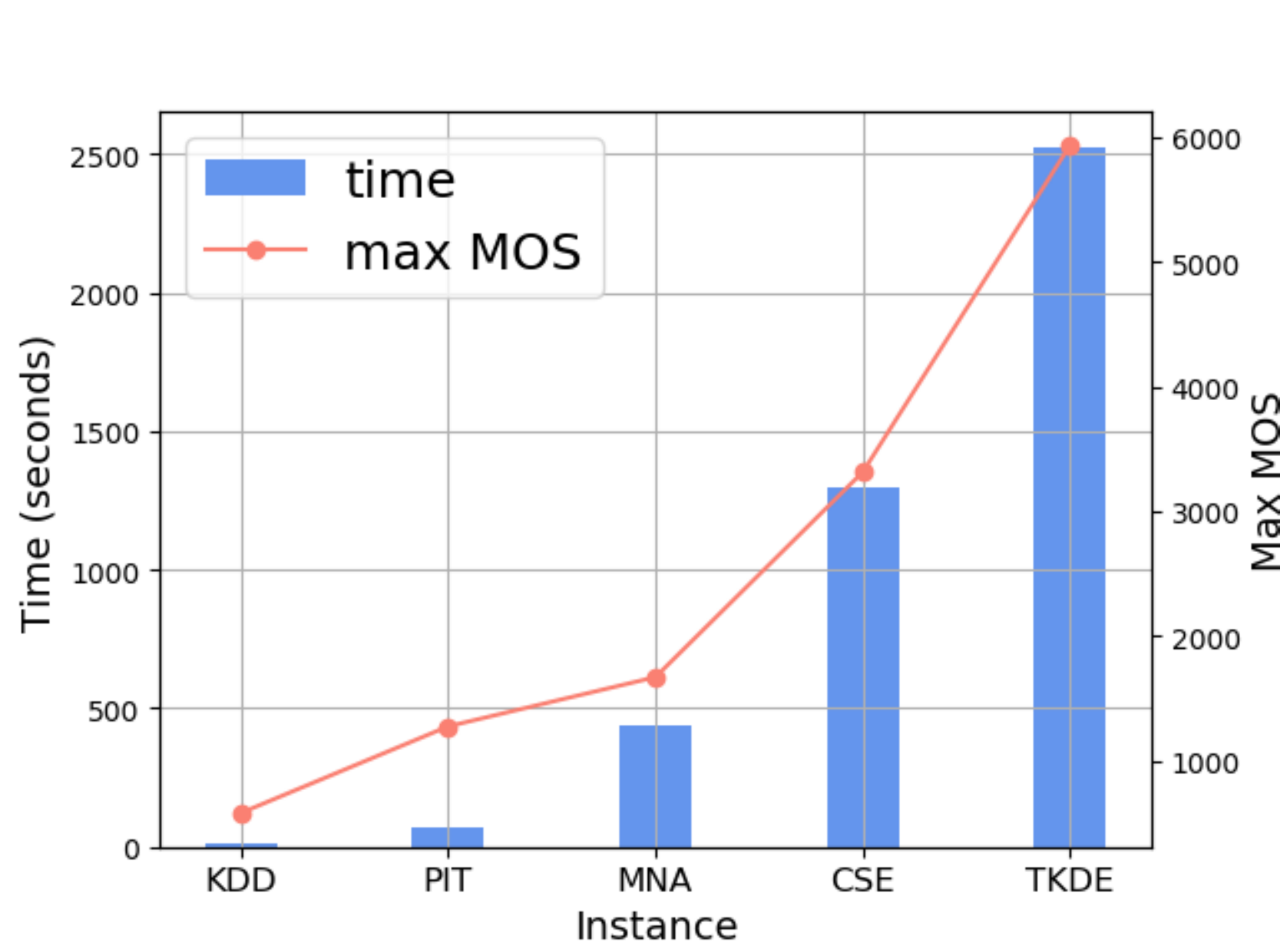}\\
    \caption{The computational time and max MOS under different network scales.}\label{fig:eff}
\end{figure}

Herein, we conduct several experiments to explore the efficiency of the algorithm under different conditions. We extract 5 different venues and list their basic information in Table~\ref{tab:eff}. The full venue names are Proceedings of the 17th ACM SIGKDD International Conference on Knowledge Discovery and Data Mining (KDD), Problems of Information Transmission (PIT), Mobile Networks and Applications (MNA), Computing in Science and Engineering (CSE), and IEEE Transactions on Knowledge and Data Engineering (TKDE). Other query conditions are the same as Case 2 in the first part of Section~\ref{sec:exp}. We use ``C. Faloutsos - H. Tong - diversified", ``V. V. Zyablov - M. Handlery - tailbiting", ``F. Xia - G. Wu - authentication", ``K. Hinsen - G. K. Thiruvathukal - version" and ``J. Tang - J. Li - recommendation" as the start motifs, respectively. The result of the experiment is shown in Fig. \ref{fig:eff}.

\begin{table}[h]
    \centering
    \caption{The basic information of different venues.}
    \resizebox{\linewidth}{!}{
        \begin{tabular}{|c|c|c|c|c|c|}
            \hline
            & KDD & PIT & MNA & CSE & TKDE\\
            \hline
            published papers & 178 & 447 & 740 & 1,338 & 2,601\\
            \hline
            edge number & 8,760 & 16,280 & 34,860 & 38,896 & 91,801\\
            \hline
            author number & 535 & 381 & 2,002 & 2,335 & 4,999\\
            \hline
            term number & 721 & 1,224 & 1,710 & 2,962 & 3,943\\
            \hline
        \end{tabular}
    }
    \label{tab:eff}
\end{table}

\begin{figure}
    \centering
    \includegraphics[width=0.8\linewidth]{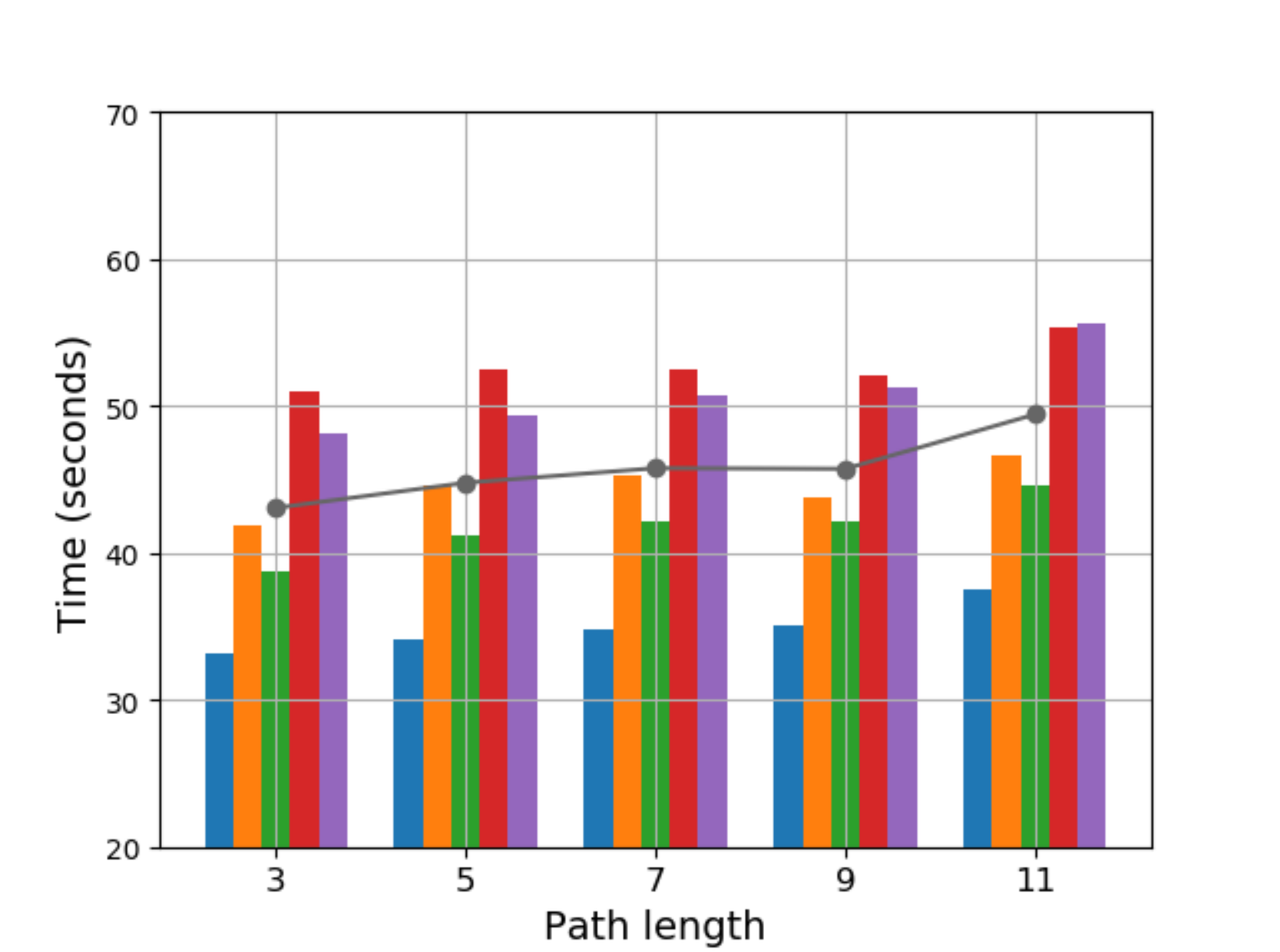}\\
    \caption{The time consuming and max MOS with the increasing length of metapaths.}\label{fig:path_len}
\end{figure}

It can be seen that the computational time grows with the increase of edge numbers. Meanwhile, we also randomly choose 5 motifs as the start motifs in each network. Then we compute the average running time with different lengths of symmetric paths. The result is shown in Figure~\ref{fig:path_len}, wherein there exist five groups. Each group contains five bars, and each bar corresponds to a meta path with a certain length. The line shows the average computational time of each group. All of metapaths start with ``author" and end with ``author". That is, a metapath with length of 3 is ``author-term-author" and with length of 5 is ``author-term-author-term-author", and so on. Interestingly, with the increasing length of meta paths, the computational time grows with a slight ascending trend. This indicates that the computational time is robust to the length of meta paths.
Combining with the basic information of networks, the computational time grows with the increase of average network degree.
Variations in computational times among different meta paths are observed, which is due to different query conditions: the more specific the query conditions are, the less time the algorithm consumes. A fuzzy query condition will lead to more searching results in the candidate set, which will lead to longer computational time. In general, the computational time of the proposed algorithm mainly depends on the average network degree and the query conditions.

\begin{figure*}[!htp]
    \centering
    \subfigure[KDD]{
        \includegraphics[width=0.18\textwidth]{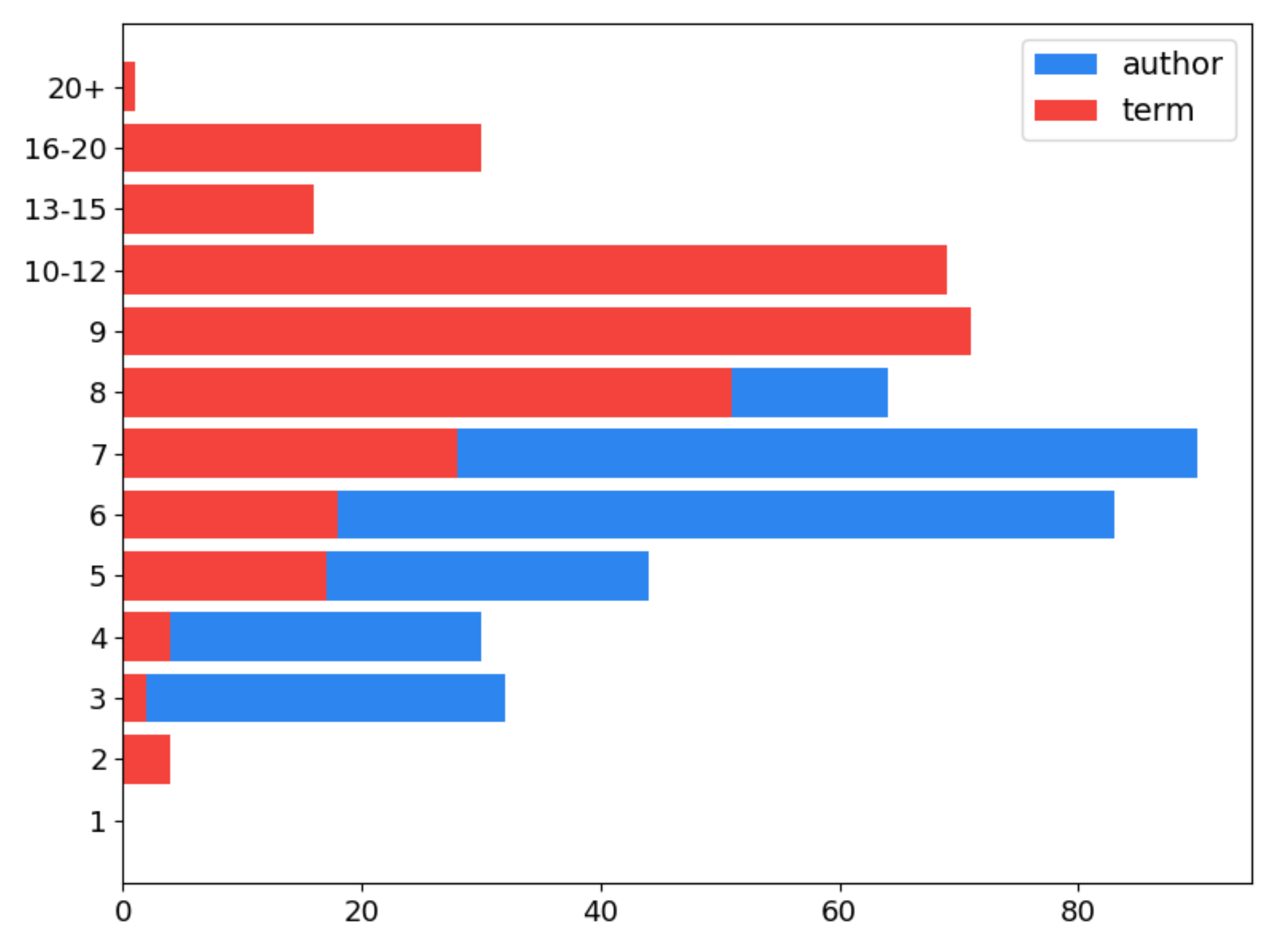}
        \label{fig:kdd}
    }
    \subfigure[PIT]{
        \includegraphics[width=0.18\textwidth]{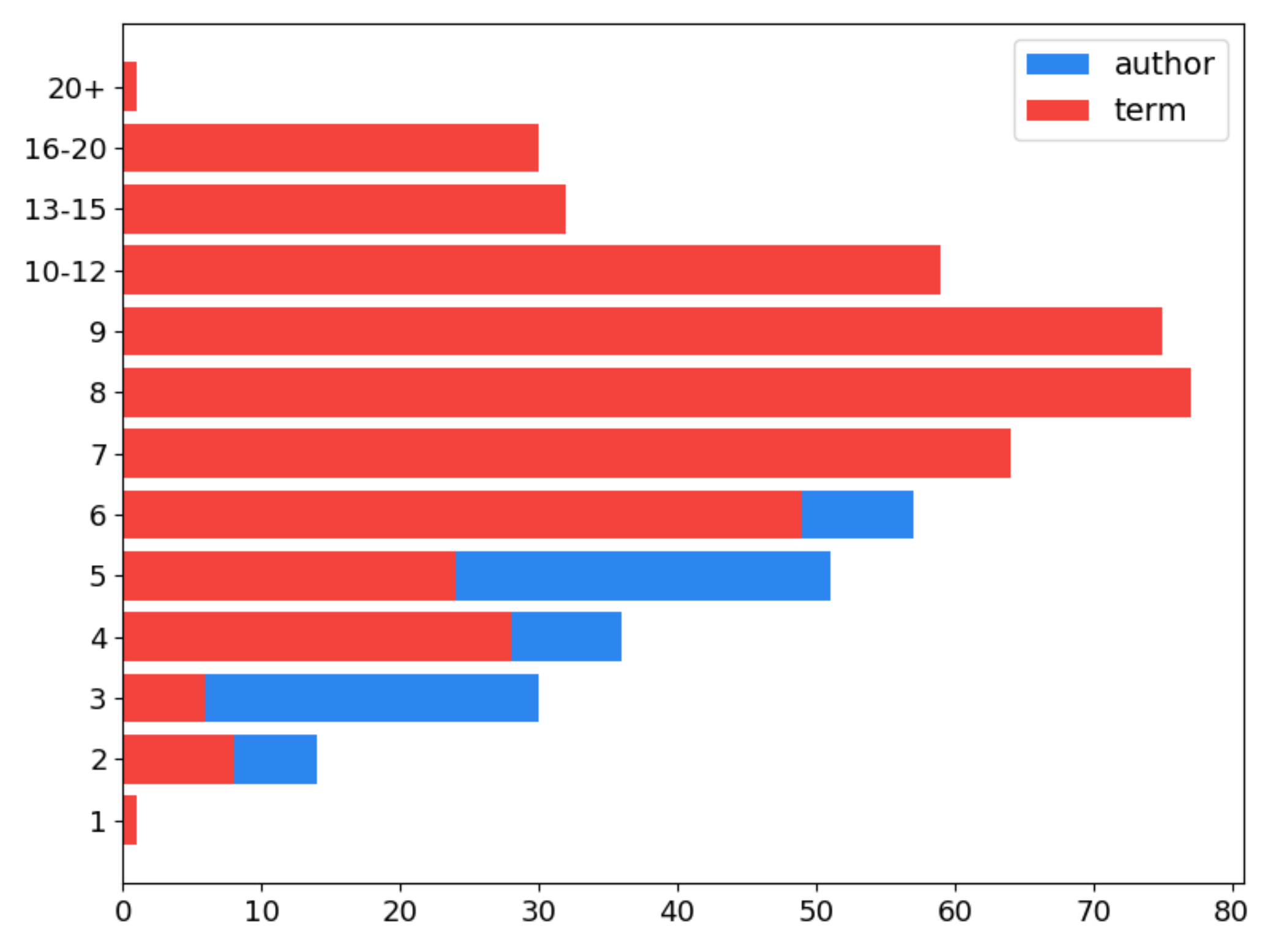}
        \label{fig:pit}
    }
    \subfigure[MNA]{
        \includegraphics[width=0.18\textwidth]{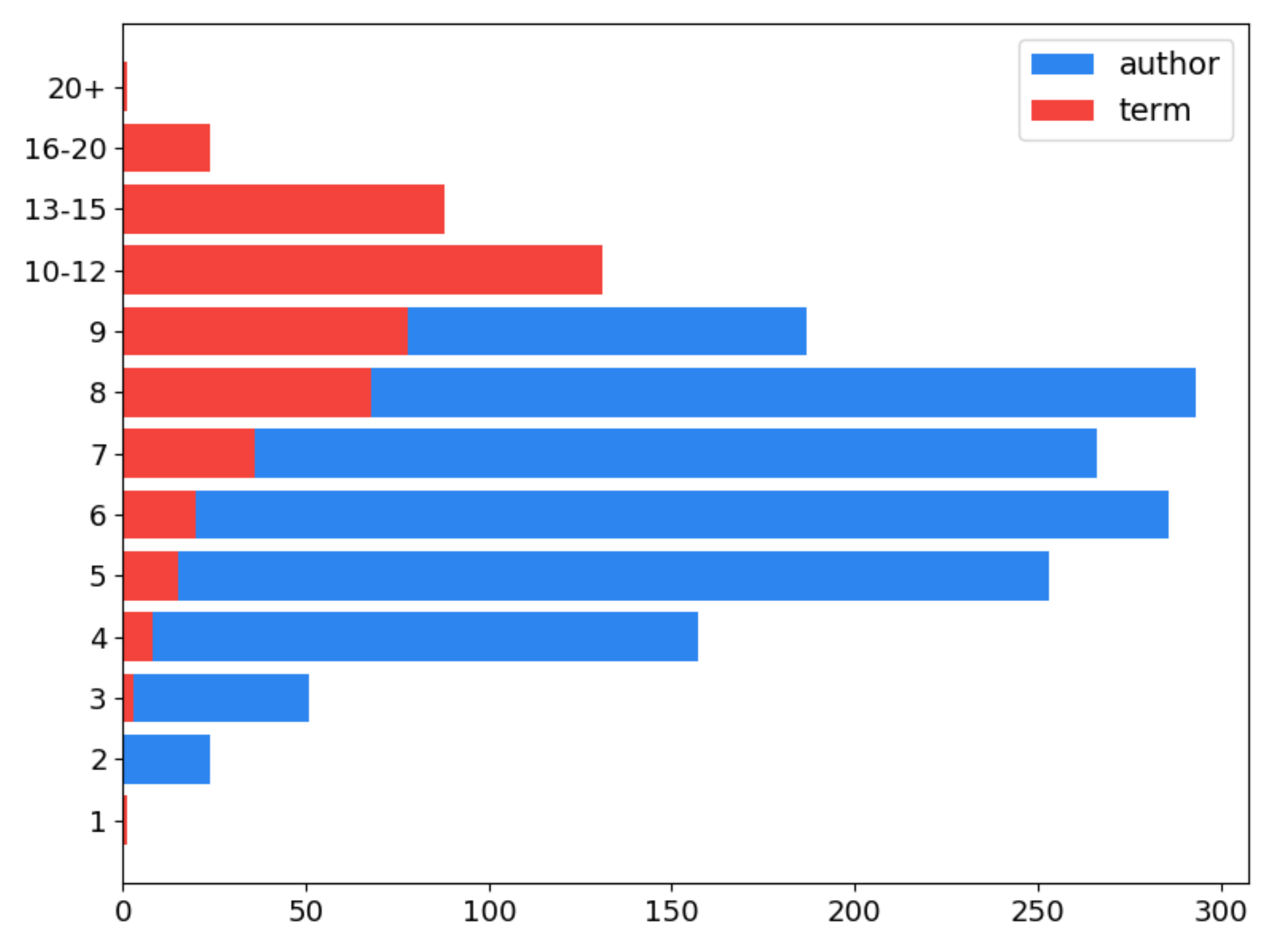}
        \label{fig:mna}
    }
    \subfigure[CSE]{
        \includegraphics[width=0.18\textwidth]{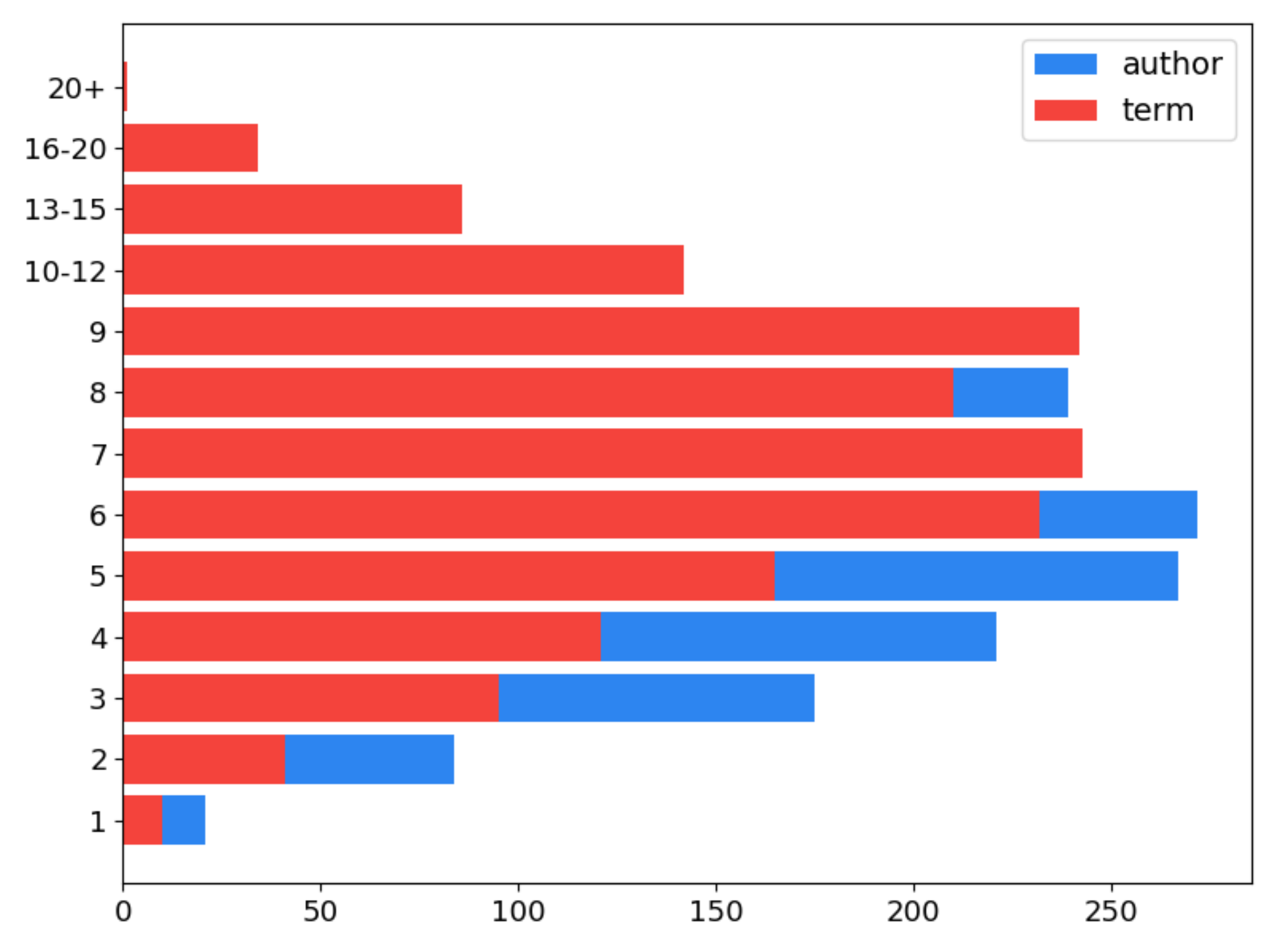}
        \label{fig:cse}
    }
    \subfigure[TKDE]{
        \includegraphics[width=0.18\textwidth]{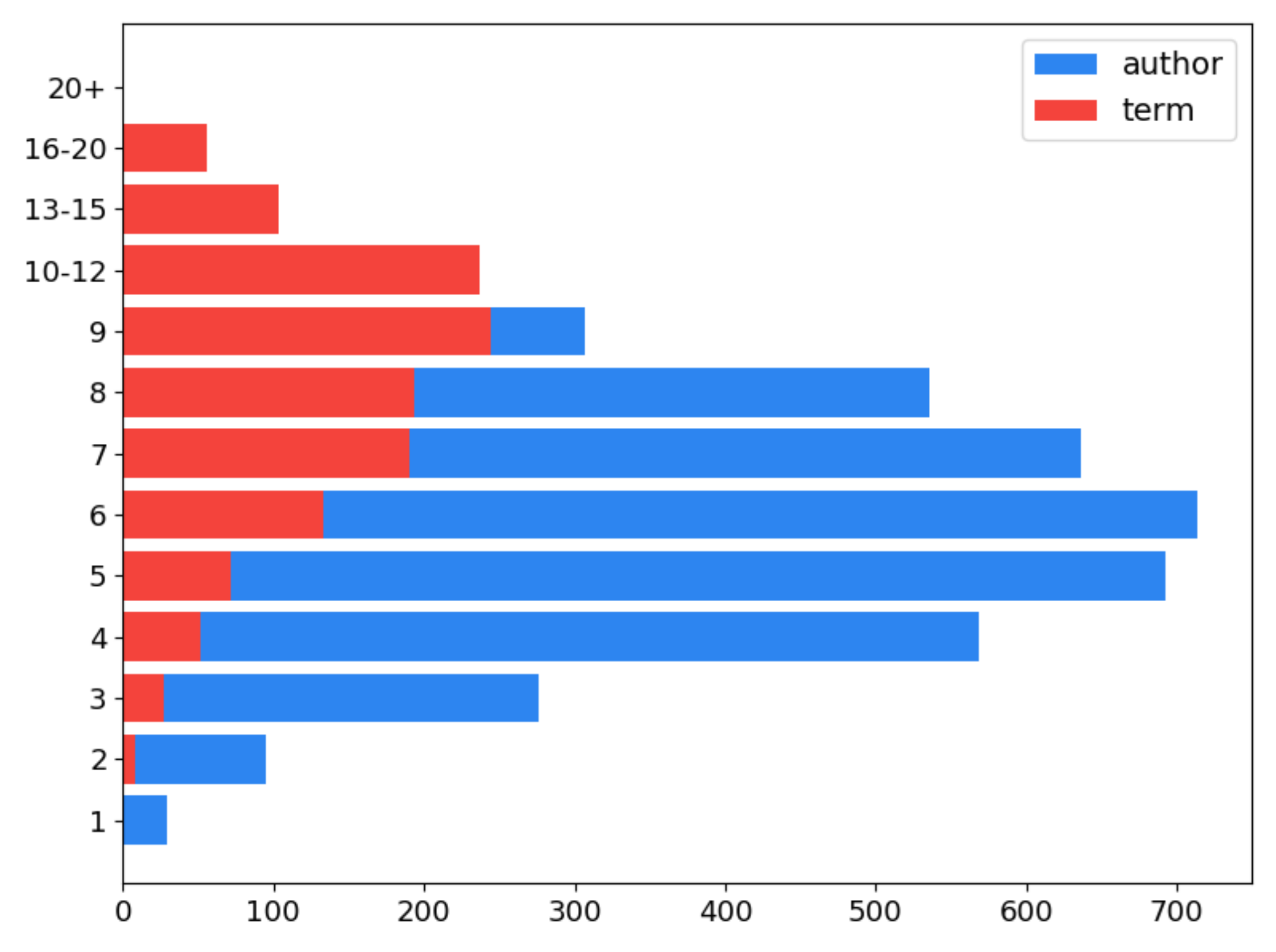}
        \label{fig:tkde}
    }
    \caption{The distribution of node degree in KDD, PIT, MNA, CSE, and TKDE.}
    \label{fig:5statistics}
\end{figure*}

Since different scales of networks will have an impact on computational time, we specifically discuss the relationships between node degree distribution and computational time. The statistics of node degree distribution is shown in Figure~\ref{fig:5statistics}. It can be seen that the node degree distribution influences on the computational time. When the networks are in smaller scales (KDD, PIT, MNA), degree distribution has little impact on computational time. CSE and TKDE are in larger scales, while the computational time of CSE is much less than that of TKDE. This is because the author node degree in CSE network is much less, meanwhile the peak node of term node degree appears earlier than that of TKDE.

\subsection{Candidate Set}

In this work, we generate the candidate set based on the reference set and the given meta path set. The motifs in the candidate set are relevant to the query motif and the motifs in the reference set.
The outlier motifs refer to the irrelevant ones according to ground truth that appear relevant in the query results. The generation of the candidate set can collect query-relevant motifs because we use symmetric meta paths. The symmetric meta path we employed can reflect the similarity between the two connected nodes. Therefore, the motifs we generated in the candidate set are generally relevant to the query motif or the motifs in the reference set.

The number of meta paths can reflect the irrelevant relationships between outlier motifs and the query motif or the motifs in the reference set. If two motifs own a large number of meta paths, then the two motifs are of closer properties as well as structural similarities. Therefore, the outlier motifs are generally irrelevant to the query motif or the motifs in the reference set.

\subsection{Meta Path}

\begin{figure}[!htp]
    \centering
    \subfigure[Terms]{
        \includegraphics[width=0.2\textwidth]{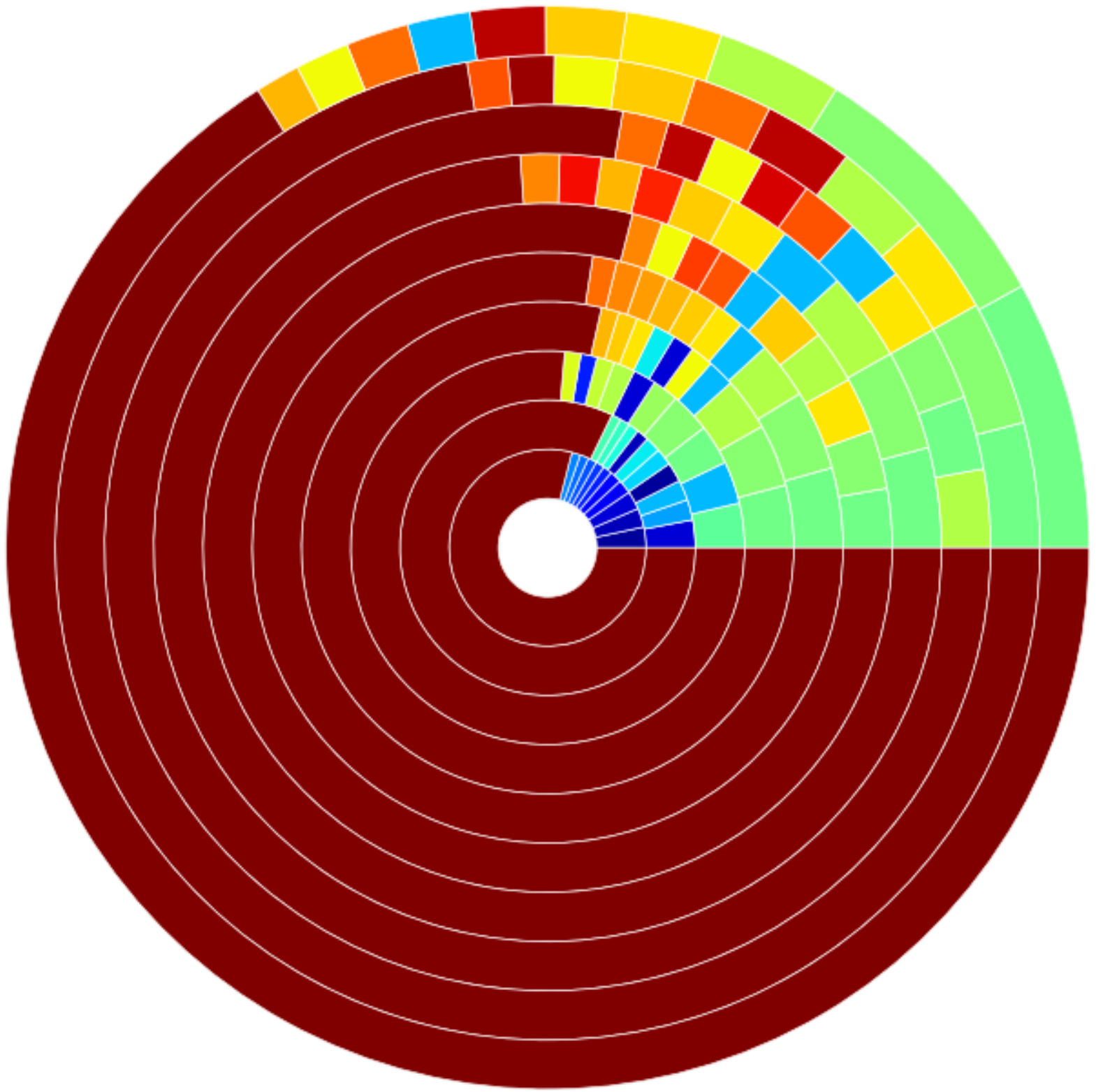}
        \label{fig:mp_k}
    }
    \subfigure[Authors]{
        \includegraphics[width=0.2\textwidth]{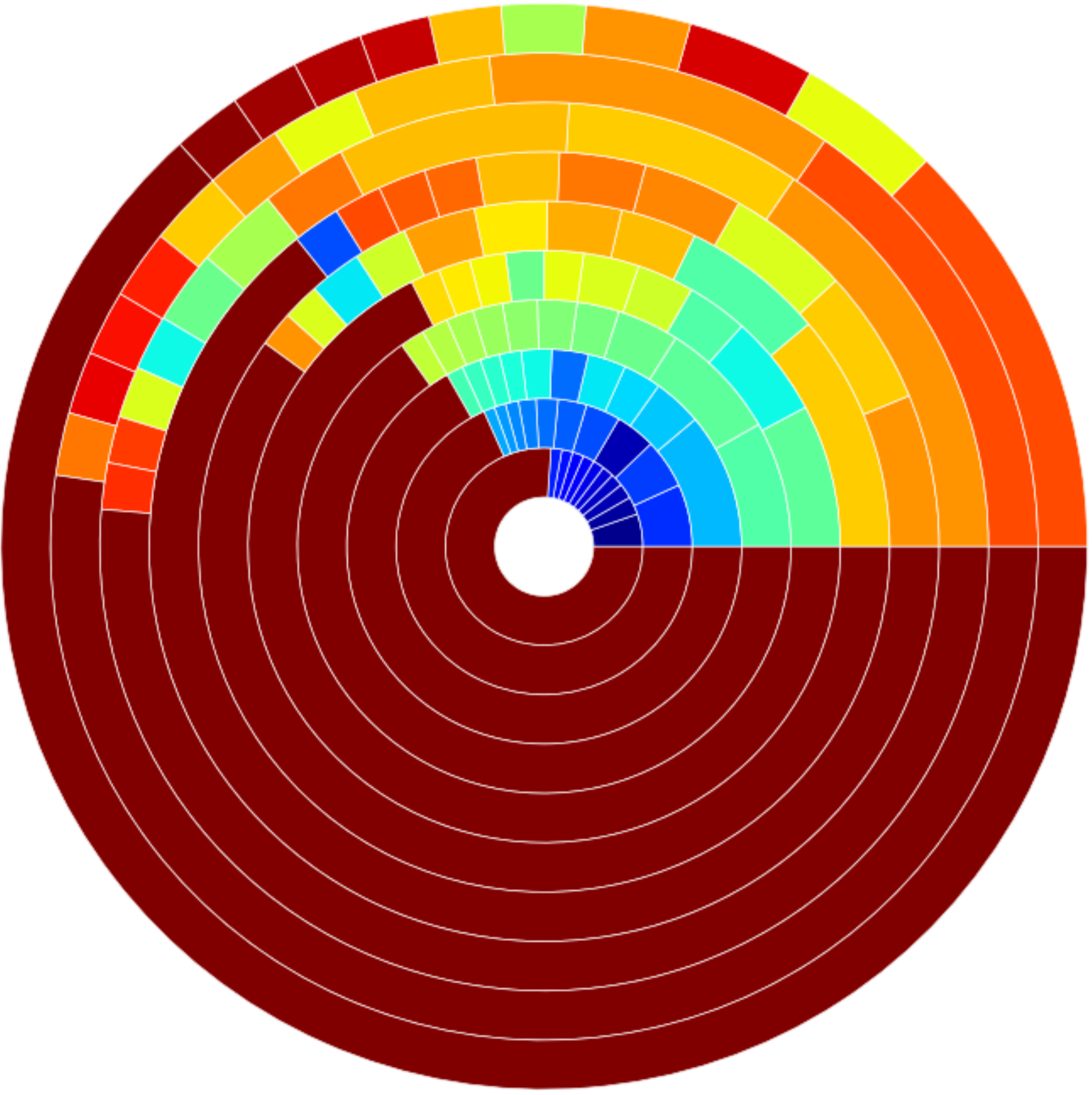}
        \label{fig:mp_a}
    }
    \caption{The distribution of nodes in outlier motifs. \ref{fig:mp_k} the distribution of ``term". \ref{fig:mp_a} the distribution of ``authors".}
    \label{fig:mp}
\end{figure}

Herein, we discuss the validness of the meta path. Although we can use different meta paths for calculating MOS, the important problem is which type of paths should we use to get a better performance? In heterogeneous information networks, different meta paths have different semantic meanings, like ``author - paper - author" means co-authorship between two authors. As a result, they will show different properties. We present the different paths for computing, which leads to different distributions in the result. Moreover, it also illustrates that more meta paths will get the exact relative order. But we cannot use excessive meta paths in the set for computing. In real life, different meta paths have different semantics. According to the meaning we are concerned about, we use different meta paths to find motifs and calculate the similarity between motifs. Herein, we implement a contrast experiment, and the experimental results are shown in Figure~\ref{fig:mp}. We also use ``F. Xia-G. Wu-authentication" as a start motif. Meanwhile, we only use one meta path (i.e., ``author-term-author") to detect outlier motifs. We achieve totally different detecting results. Comparing with two meta paths (i.e., ``author-term-author" and ``term-author-term"), one meta path gains more fuzzy searching results. The reason behind may be that the searching results are generally influenced by a certain node within the start motif when employing only one meta path. Apparently, more meta path can improve the searching accuracy but decline the consuming time in the meantime.

There are also some works on the reliability of meta path \cite{shams2018reliable}. However, the conclusion is not comprehensive for all types of heterogeneous networks. We also need to find a common standard for various heterogeneous networks. On the other hand, an effective method is also vital for finding fitting paths in calculating similarity.
\begin{table}[h]
    \centering
    \caption{The top 10 outlier motifs and top 10 most similar motifs detected based on the query motif ``Feng Xia (author) - G. Wu (author) - authentication (term)" and PathSim similarity.}
    \resizebox{\linewidth}{!}{
        \begin{tabular}{|c|c|c|c|}
            \hline
            Author 1 & Author 2 & Term & PathSim\\
            \hline
            X. Geng&Y. Huang&defending&41.397\\
            X. Geng&Y. Huang&against&41.397\\
            X. Geng&Y. Huang&challenge&41.397\\
            X. Geng&Y. Huang&ddos&41.397\\
            X. Geng&A. B. Whinston&defending&41.397\\
            X. Geng&A. B. Whinston&against&41.397\\
            X. Geng&A. B. Whinston&challenge&41.397\\
            X. Geng&A. B. Whinston&ddos&41.397\\
            Y. Huang&A. B. Whinston&defending&41.397\\
            Y. Huang&A. B. Whinston&against&41.397\\
            \(\vdots\) & \(\vdots\) & \(\vdots\) & \(\vdots\)\\
            H. Chen&Y. Zhang&wimax&1460.127\\
            Y. Zhang&H. Chen&resilient&1460.299\\
            Y. Zhang&H. Chen&optimized&1469.887\\
            V. C. Leung&Y. Zhang&controlled&1470.980\\
            H. Chen&Y. Zhang&inter-carrier&1471.115\\
            V. C. Leung&Y. Zhang&fusion&1479.168\\
            Y. Zhang&V. C. Leung&802.16&1482.189\\
            H. Chen&Y. Zhang&qos-aware&1485.277\\
            H. Chen&Y. Zhang&uplink&1485.788\\
            H. Chen&Y. Zhang&802.16&1527.335\\
            \hline
        \end{tabular}
    }
    \label{tab:ex1-path}
\end{table}
\subsection{Similarities}
We discuss the influences of finding outlier motifs according to different similarities. Herein, we detect the outlier motifs based on PathSim and CosSim, respectively. The results of PathSim are shown in Table~\ref{tab:ex1-path}. Comparing with MOS, we can see that the most similar 10 motifs in Table~\ref{tab:ex1-path} are completely different from those in Table~\ref{tab:ex1}. However, the top 10 outlier motifs are exactly the same, even with the same ranking order. Moreover, each motif listed in the top 10 outlier motifs owns the same similarity distribution. That is, though different ways calculate the similarity values, outlier motifs with the same MOS value are detected in same orders. For example, in Table~\ref{tab:ex1-path}, all of the 10 outlier motifs are with the same PathSim equal to 41.397. In Table~\ref{tab:ex1}, these 10 outlier motifs own the same MOS values of 58.867. In Table~\ref{tab:ex1-cos}, the CosSim values of these outlier motifs are all 42.038.
\begin{table}
    \centering
    \caption{The top 10 outlier motifs and top 10 most similar motifs detected based on the query motif ``Feng Xia (author) - G. Wu (author) - authentication (term)" and the CosSim similarity.}
    \resizebox{\linewidth}{!}{
        \begin{tabular}{|c|c|c|c|}
            \hline
            Author 1 & Author 2 & Term & CosSim\\
            \hline
            X. Geng&Y. Huang&defending&42.038\\
            X. Geng&Y. Huang&against&42.038\\
            X. Geng&Y. Huang&challenge&42.038\\
            X. Geng&Y. Huang&ddos&42.038\\
            X. Geng&A. B. Whinston&defending&42.038\\
            X. Geng&A. B. Whinston&against&42.038\\
            X. Geng&A. B. Whinston&challenge&42.038\\
            X. Geng&A. B. Whinston&ddos&42.038\\
            Y. Huang&A. B. Whinston&defending&42.038\\
            Y. Huang&A. B. Whinston&against&42.038\\
            \(\vdots\) & \(\vdots\) & \(\vdots\) & \(\vdots\)\\
            H. Chen&H. Hu&ad-hoc&1389.960\\
            H. Chen&C. Wang&ad-hoc&1404.822\\
            H. Chen&Y. Huang&ad-hoc&1411.361\\
            H. Chen&C. Yeh&ad-hoc&1417.390\\
            H. Chen&T. Wang&ad-hoc&1417.390\\
            H. Chen&K. Conner&ad-hoc&1421.122\\
            H. Chen&T. Rasheed&ad-hoc&1421.122\\
            H. Chen&D. Meddour&ad-hoc&1429.639\\
            S. Mao&V. C. Leung&ad-hoc&1439.300\\
            H. Chen&Y. Zhang&ad-hoc&1496.971\\
            \hline
        \end{tabular}
    }
    \label{tab:ex1-cos}
\end{table}

CosSim reflects the similarity between two nodes from the perspective of the cosine value between two vector angles in a vector space. Herein, we list the detecting results in Table~\ref{tab:ex1-cos}. Comparing with NetOut and PathSim, CosSim yields identical detecting results. There exists another interesting phenomenon that the top 10 outlier motifs have the same similarity values, including MOS, PathSim, and CosSim. It seems that different similarities barely influence on the outlier motif detection results. This phenomenon indicates that our proposed algorithm performs well in detecting outlier motifs. Furthermore, the proposed algorithm is insensitive to different similarities when detecting outlier motifs.

As for the most similar motifs in our result \cref{tab:ex1-path,tab:ex1-cos}, there exist some differences. None of the top 10 most similar motifs are the same in the three tables. Some of the nodes or edges within the motifs may be the same, but the whole motifs are not.
The top 10 most similar motifs are different may be caused by different similarity metrics.

\begin{table}[h]
  \centering
  \caption{The top 10 outliers and top 10 most similar to given motif set using PathSim. The beginning motif is ``J. Tang (author) - J. Li (author) - recommendation (term)".}
  \resizebox{\linewidth}{!}{
    \begin{tabular}{|c|c|c|c|}
      \hline
      Author 1 & Author 2 & Term & PathSim\\
      \hline
      W. Li&D. yeung&mild&16.358\\
      W. Li&D. yeung&multiple-instance&16.358\\
      M. G. Hwang&C. Choi&sense&39.134\\
      M. G. Hwang&P. Kim&sense&39.134\\
      C. Choi&P. Kim&sense&39.134\\
      M. G. Hwang&C. Choi&enrichment&39.938\\
      M. G. Hwang&P. Kim&enrichment&39.938\\
      C. Choi&P. Kim&enrichment&39.938\\
      A. N. Zincir-Heywood&M. I. Heywood&object-orientated&46.706\\
      A. N. Zincir-Heywood&C. R. Chatwin&object-orientated&46.706\\
      \(\vdots\) & \(\vdots\) & \(\vdots\) & \(\vdots\)\\
      X. Lin&Y. Tao&multivalued&5651.753\\
      X. Lin&Y. Tao&borda&5651.753\\
      Y. Tao&X. Lin&anonymous&5680.435\\
      Y. Tao&X. Lin&dimensional&5689.031\\
      Y. Tao&X. Lin&skylines&5690.555\\
      Y. Tao&X. Lin&threshold-based&5699.910\\
      Y. Tao&X. Lin&k&5699.910\\
      Y. Tao&X. Lin&existentially&5700.076\\
      Y. Tao&X. Lin&extents&5736.802\\
      Y. Tao&X. Lin&medium&5752.686\\
      \hline
    \end{tabular}}
  \label{tab:ex2-path}
\end{table}
\begin{table}
    \centering
    \caption{The top 10 outliers and top 10 most similar to given motif set using CosSim. The beginning motif is ``J. Tang (author) - J. Li (author) - recommendation (term)".}
    \resizebox{\linewidth}{!}{
        \begin{tabular}{|c|c|c|c|}
            \hline
            Author 1 & Author 2 & Term & CosSim\\
            \hline
            W. Li&D. yeung&mild&18.356\\
            W. Li&D. yeung&multiple-instance&18.356\\
            M. G. Hwang&C. Choi&sense&40.436\\
            M. G. Hwang&P. Kim&sense&40.436\\
            C. Choi&P. Kim&sense&40.436\\
            M. G. Hwang&C. Choi&enrichment&41.740\\
            M. G. Hwang&P. Kim&enrichment&41.740\\
            C. Choi&P. Kim&enrichment&41.740\\
            A. N. Zincir-Heywood&M. I. Heywood&object-orientated&48.951\\
            A. N. Zincir-Heywood&C. R. Chatwin&object-orientated&48.951\\
            \(\vdots\) & \(\vdots\) & \(\vdots\) & \(\vdots\)\\
            X. Lian&Lei Chen&cost&4152.757\\
            X. Lian&Lei Chen&range&4156.470\\
            Y. Tao&X. Lin&location-based&4158.846\\
            Y. Tao&X. Lin&medium&4168.364\\
            Y. Tao&X. Lin&neighbor&4179.917\\
            Y. Tao&X. Lin&nearest&4214.662\\
            X. Lian&Lei Chen&neighbor&4232.639\\
            Y. Tao&X. Lin&reverse&4236.367\\
            X. Lian&Lei Chen&nearest&4265.022\\
            Lei Chen&X. Lian&reverse&4278.532\\
            \hline
    \end{tabular}}
    \label{tab:ex2-cos}
\end{table}
To verify the similarity's influence on the detection results, we implement another two experiments for Case 2 in an academic network, which starts from motif ``J. Tang (author) - J. Li (author) - recommendation (term)". We implement our experiments with PathSim and CosSim, respectively. Table~\ref{tab:ex2-path} shows the detecting results of PathSim, and Table~\ref{tab:ex2-cos} shows that of CosSim. In Table~\ref{tab:ex2-path}, the list of the top 10 outlier motifs is similar but not the same as the experimental results using MOS. Generally, the top 8 outlier motifs detected based on PathSim are identical as that of MOS. The other 2 outlier motifs are quite different from that in Table~\ref{tab:ex2}. For detecting results in Table~\ref{tab:ex2-cos}, there also exist the same top 8 outlier motifs.

Similar to Case 1, the similarity values of the top 10 outliers appear with the same ranking orders. Among all the detecting results based on these three similarity metrics, the first top 2 outlier motifs own the same value. This illustrates that the proposed algorithm is a general method that will not be affected by different similarities. In other words, our proposed algorithm can detect outlier motifs with a stable performance. As for the top 10 most similar motifs, few nodes or edges of Table~\ref{tab:ex2-path} or Table~\ref{tab:ex2-cos} are the same comparing with MOS. This means that different similarities lead to different detecting results, but the proposed algorithm reduces or eliminates such cases. Therefore, the proposed method can achieve relative stable detecting results.

\section{Conclusion}
\label{sec:con}
In this work, we have examined outlier motifs, an interesting and critical issue in social computing. We have proposed an efficient algorithm for finding outlier motifs in heterogeneous information networks. By exploring the user's query and constrained conditions (i.e. human behaviour), we calculate MOS of each motif in the candidate motif set and sort the MOS values in the ascending order. We set the standard for MOS in query, which is the structure of a motif similar to a reference motif set. The proposed algorithm contains four steps: obtaining motifs meeting query conditions, counting meta paths between nodes, calculating MOS between a motif in the candidate motif set and the whole reference motif set, and ordering MOS of each motif in the candidate motif set. We verify our algorithm on two information networks from the real-world academic network and discuss the experimental results of both networks. Interesting outlier motifs are also found in these networks. Furthermore, we also discuss the details about how to choose the candidate set, meta path, etc. Our work sheds light on finding interesting outlier motifs in large-scale heterogeneous networks and provides a new way of outlier detection for human behavior patterns.

\bibliographystyle{ACM-Reference-Format}
\bibliography{reference}

\end{document}